\documentclass[aps,prb,reprint,twocolumn,floatfix,superscriptaddress,amsmath,amssymb,amsfonts,nofootinbib,balancelastpage,longbibliography,superscriptaddress]{revtex4-2}

\usepackage{url}
\usepackage{bm}
\usepackage{graphicx}
\usepackage{amsmath}
\usepackage{amstext}
\usepackage{amssymb}
\usepackage{amsfonts}
\usepackage{amsbsy}
\usepackage{verbatim}
\usepackage{color}
\usepackage{floatrow}
\usepackage{float}
\usepackage{tikz}
\usepackage{gensymb}
\usepackage{textcomp}
\usepackage{enumitem}
\usepackage[version=3]{mhchem}
\usepackage{afterpage}
\usepackage{pbox}
\usepackage{dsfont}
\usepackage{siunitx}
\usepackage{stackengine,wasysym,scalerel}
\usepackage{latexsym}
\usepackage{cancel}
\usepackage{comment}
\usepackage{array, multirow, bigdelim, makecell, booktabs}
\usepackage[misc]{ifsym}
\usepackage{sidecap}
\usepackage{color}
\usepackage{pifont}

\usepackage[normalem]{ulem}

\definecolor{darkblue}{HTML}{004D6B}
\definecolor{darkred}{HTML}{8c1515}
\definecolor{darkgreen}{HTML}{006400}
\usepackage[colorlinks=true, urlcolor=blue, linkcolor=blue, citecolor=blue, pdftex]{hyperref}
\usepackage[capitalise]{cleveref}

\usepackage{xcolor}

\renewcommand{\phi}{\varphi}
\renewcommand{\epsilon}{\varepsilon}

\newcommand{\eqnref}[1]{Eq.~(\ref{#1})}

\newcommand{\appref}[1]{Appendix~\ref{#1}}

\newcommand{\A}{{\mathcal{A}}}
\newcommand{\Acsm}{{\hat{\mathcal{A}}}}

\newcommand{\ket}[1]{\left|#1\right\rangle}

\usepackage{tikz}
\usetikzlibrary{arrows.meta, positioning}
\usepackage{mathtools}

\begin{document}

\title{Parent Hamiltonian construction of generalized Calogero--Sutherland models}

\author{Hari Borutta$^{*,\dagger}$}
\affiliation{Department of Physics, Indian Institute of Technology Madras, Chennai 600036, India}
\affiliation{W\"urzburg--Dresden Cluster of Excellence ctd.qmat, 97074 W\"urzburg and 01062 Dresden, Germany}

\author{Andreas Feuerpfeil$^{*,\ddagger}$}
\affiliation{W\"urzburg--Dresden Cluster of Excellence ctd.qmat, 97074 W\"urzburg and 01062 Dresden, Germany}
\affiliation{Institut f\"ur Theoretische Physik und Astrophysik, Julius-Maximilians-Universit\"at W\"urzburg, Am Hubland, Campus S\"ud, W\"urzburg 97074, Germany}

\author{Yasir Iqbal$^{\S}$}
\affiliation{Department of Physics, Indian Institute of Technology Madras, Chennai 600036, India}

\begingroup
\renewcommand\thefootnote{\fnsymbol{footnote}}
\footnotetext[1]{These authors contributed equally to this work.}
\footnotetext[2]{\href{mailto:haripbj@physics.iitm.ac.in}{haripbj@physics.iitm.ac.in}}
\footnotetext[3]{\href{mailto:andreas.feuerpfeil@uni-wuerzburg.de}{andreas.feuerpfeil@uni-wuerzburg.de}}
\footnotetext[4]{\href{mailto:yiqbal@physics.iitm.ac.in}{yiqbal@physics.iitm.ac.in}}
\endgroup

\date{\today}

\begin{abstract}
The Calogero--Sutherland model is a paradigmatic integrable system describing one-dimensional non-relativistic particles with inverse-square-type interactions. At interaction strength $\lambda=2$, the CSM exhibits a deep connection to anyon physics, featuring the Laughlin--Jastrow polynomial as its exact ground state. Motivated by this structure, we develop a general reverse-engineering construction of positive semidefinite continuum parent Hamiltonians for trial states admitting a rational conformal field theory description with central charge $c<1$. By leveraging the null-vector structure of the underlying primary fields and the associated Belavin--Polyakov--Zamolodchikov equations, we derive corresponding many-body annihilation operators. We then apply this construction explicitly to the Moore--Read and $k=3$ Read--Rezayi states---relating to Ising and Fibonacci anyons, respectively---obtaining continuum Hamiltonians for which these Jack-polynomial states, viewed as one-dimensional chiral spin liquid wave functions, are exact zero modes. We emphasize, however, that our construction does not by itself establish ground-state uniqueness or determine the nature of the excitation spectrum.
\end{abstract}


\maketitle

\section{Introduction}
Framed in the context of the fractional quantum Hall effect, the notion of topological order~\cite{RevModPhys.89.041004} has evolved into a central motif of condensed matter research in recent decades. This concept has subsequently carried over to systems as diverse as superconductors, spin liquids, semiconductor-superconductor heterostructures, fractional Chern insulators and beyond. As a key consequence of topological order, the field witnessed the conception of fractional quasiparticle statistics beyond the familiar choices of bosons and fermions~\cite{greiter-24arcmp131}. Related to the anyon statistics, the excitations possess fractional quantum numbers relative to the bare particles which span the Hilbert space. In particular, this encompasses non-Abelian particles~\cite{Moore_1991}, where the quasiparticle exchange implies an overall rotation within an approximately degenerate ground state manifold. This degenerate space is topologically protected against local perturbations, and hence relevant to topological quantum computation~\cite{nayak_08}.

Initially, as suggested by the nature of irreducible braid group representations, anyons in physically realistic contexts were believed to be constrained to two spatial dimensions~\cite{khare2005fractional}. While fractons broadened the perspective toward anyons in higher dimensions~\cite{RevModPhys.96.011001}, anyons in one spatial dimension first challenged the reconciliation of fractional statistics from the viewpoint of particle exchange and a generalized exclusion principle~\cite{PhysRevLett.67.937,PhysRevLett.73.1574,PhysRevB.79.064409}, which was later extended to non-Abelian anyons in one dimension~\cite{Greiter2011,PhysRevB.100.115107}. There, the key insight to non-Abelian state counting in one dimension is an occupation-dependent exclusion principle reminiscent of the state-counting rules for electrons on the quantum Hall sphere~\cite{haldane83prl605,PhysRevB.84.045127}.

The Calogero--Sutherland Model (CSM)~\cite{calogero69jmp2191,calogero69jmp2197,sutherland71jmp246,sutherland71pra2019,sutherland72pra1372} describes a class of integrable continuum models of non-relativistic particles in one spatial dimension with inverse-square potential~\cite{10.1007/3-540-58453-6_3,ferrando2025bethe}. Unlike most Bethe-integrable models, which do not easily facilitate access to correlation functions, the CSM is rigorously accessible. Depending on the potential strength $\lambda$, the CSM relates to the Brownian motion in Wigner-Dyson random matrix ensembles with orthogonal, unitary, or symplectic symmetry. This feature renders it amenable to rigorous calculations not only of its spectrum but also of dynamical signatures, for instance through its exact one- and two-particle Green’s functions~\cite{PhysRevB.52.8729}. This exact solvability is closely tied to the fact that, for special values of $\lambda$, the excitations of the CSM can be interpreted in terms of an ideal gas of particles with fractional exclusion statistics, including fermions at $\lambda=1$ and half-fermions at $\lambda=2$.

Being a {\it sui generis} continuum quantum model of interacting particles, the CSM has nonetheless triggered enormous interest and reached high relevance in the mathematical foundation of anyonic quantum particles in one dimension. For instance, the CSM is intimately related to the Haldane--Shastry spin chain model (HSM)~\cite{haldane88prl635,shastry88prl639}, as the HSM can be reconciled as the Hamiltonian governing the spin-wave quantum fluctuations around the static equilibrium particle positions in the CSM when charge fluctuations are frozen out~\cite{polychronakos93prl2329}.  The HSM, in turn, has been generalized to $\mathrm{SU}(2)_k$ WZW models~\cite{Nielsen_2011} and non-Abelian spin chains~\cite{Greiter2011,PhysRevB.85.195149}, whose exact ground-state polynomial matches the non-Abelian chiral spin liquid series~\cite{PhysRevLett.99.097202,PhysRevB.80.104406,PhysRevLett.102.207203,PhysRevB.89.165125} featuring Ising-type spinons for $S=1$ as well as Fibonacci-type spinons for $S=3/2$. It also lays the foundation for the formulation of non-Abelian statistics in one dimension~\cite{PhysRevB.100.115107}. Any Yangian-type integrability structure, however, as exhibited by the HSM~\cite{PhysRevLett.69.2021,JCTalstra_1995}, has not yet been found for such non-Abelian spin chains.

Identifying CSM-type Hamiltonians for non-Abelian trial states is therefore of interest for several reasons. First, it might be valuable to obtain a continuum quantum model whose quasiparticle structure in the freezing limit of strongly interacting particles would relate to quantum models such as the non-Abelian spin chains. In particular, this would allow us to study the robustness of non-Abelian excitations in one dimension, and might broaden the class of models exhibiting such properties. Second, any CSM-type Hamiltonian might shed further light on the possibility of obtaining integrable models with such properties, i.e., whether there might exist an integrable model limit of an ideal gas of non-Abelian quasiparticles which, in contrast to previous attempts~\cite{PhysRevLett.98.160409}, might still allow for an embedding into a Hilbert space spanned by generic microscopic spin and charge degrees of freedom.

We emphasize that our goal is not to propose an alternative to the established lowest-Landau-level pseudopotential parent Hamiltonians for the Moore--Read and Read--Rezayi FQH states. Those Hamiltonians remain the natural framework for the two-dimensional FQH problem. The question addressed here is different. We ask whether the Calogero--Sutherland realization of the Laughlin--Jastrow polynomial admits a non-Abelian counterpart: namely, a one-dimensional continuum Hamiltonian on the circle, for which the corresponding Jack-polynomial wave function is an exact zero mode. From this perspective, our construction is complementary to the FQH parent-Hamiltonian approach, in the same way that the ordinary CSM provides a one-dimensional continuum realization of the Laughlin--Jastrow polynomial rather than a replacement for its two-dimensional pseudopotential parent Hamiltonian. The natural setting of the present work is therefore the Calogero--Sutherland/Haldane--Shastry framework and its possible extension to non-Abelian spin-chain physics.

The article is organized as follows. In Sec.~\ref{sec:CSM}, we review the analysis of the CSM to the extent necessary to appreciate its Laughlin-type ground state wave function as well as its Hamiltonian decomposition into a positive semidefinite bilinear of operators whose kernel contains the Laughlin polynomial. In Sec.~\ref{sec:csm_prescription} we describe the reverse-engineering construction to obtain parent Hamiltonians for the FQH-inspired Jack-polynomial wave functions that can be written as correlators of rational conformal field theories (RCFTs)~\cite{1984NuPhB.241..333B,ginsparg_1988,DiFrancesco_1997} with central charge $c<1$, which we then apply to the non-Abelian $k=2$ Moore--Read state and the $k=3$ Read--Rezayi state in Secs.~\ref{sec:pfaffian} and~\ref{sec:rr}, respectively. We conclude in Sec.~\ref{sec:conclusion} with a discussion of the extent to which our construction of CSM-type continuum models promises further insights into the nature of non-Abelian anyon models in 1D.

\section{Review of the Calogero--Sutherland Model}\label{sec:CSM}
The CSM~\cite{calogero69jmp2191,calogero69jmp2197,sutherland71jmp246,sutherland71pra2019,sutherland72pra1372} is a one-dimensional integrable model in which $N$ particles move along a straight line and interact via a two-particle long-range inverse sine-square-type potential\footnote{In~\cite{calogero69jmp2191,calogero69jmp2197,sutherland71jmp246,sutherland71pra2019,sutherland72pra1372}, the interaction potential is of the type $1/x_{ij}^2$, where $x_{ij} = x_i - x_j$, between particles $i$ and $j$. Additionally, there is a mutual harmonic interaction or an external harmonic well potential to ensure that the particles are bound within a finite region of space. Here, however, we consider the inverse sine-square potential, as is customary in much of the CSM literature.}. In the following, we consider periodic boundary conditions, for which we can represent the coordinates $x_j$ as complex numbers $z_j=e^{ix_j}$ on the unit circle. The CSM can then be recast as
\begin{equation} \label{eq:CSM_hamiltonian}
  \mathcal{H_{\rm CSM}}  
  =\sum_{i} \left(z_i\partial_i\right)^2
  -\lambda(\lambda-1)\sum_{i,j\atop i\neq j} \frac{z_iz_j}{(z_i-z_j)^2}\,,
\end{equation}
where $\lambda$ is the interaction strength and $\partial_i\equiv \frac{\partial}{\partial z_i}$.

The ground state of \eqnref{eq:CSM_hamiltonian} is given by
\begin{equation} \label{eq:CSM_GS}
    \Psi_{\text{GS}}(z) = \prod_{i<j} (z_i-z_j)^{\lambda} \prod_{l} z_l^{-\lambda(N-1)/2}\,.
\end{equation}
Up to the phase factor $\prod_l z_l^{-\lambda (N-1)/2}$, which has unit modulus on the circle, its polynomial part is the Laughlin-Jastrow wave function
\begin{equation}\label{eq:LJ_factor}
    \Psi_{\text{LJ}}^\lambda(z)=\prod_{i<j} (z_i-z_j)^{\lambda}\,,
\end{equation}
which models the fractional quantum Hall (FQH) effect of bosons (fermions) for even (odd) $\lambda$ at filling fraction $\nu=1/\lambda$~\cite{PhysRevLett.50.1395}. The Laughlin state hosts Abelian anyons, a genuinely two-dimensional notion; the one-dimensional counterpart is the fractional exclusion statistics obeyed by the CSM excitations, and the common Jastrow structure of the two ground states underlies this correspondence. As shown by Moore and Read~\cite{Moore_1991}, the Laughlin wave function and more general FQH wave functions have an intricate connection to conformal field theory (CFT), as they can be written as conformal blocks of a neutral CFT and a chiral boson CFT. For example, the Laughlin wave function can be expressed as the correlator of the vertex operators $:\!e^{i\sqrt{\lambda}\phi(z)}\!:$ of a chiral boson $\phi(z)$:
\begin{equation} \label{eq:LJ_correlator}
  \left\langle \!\prod_{i}\!:\!e^{i\sqrt{\lambda}\phi(z_i)}\!:\! e^{-i N \sqrt{\lambda} \phi(\infty)} \!\right\rangle=\prod_{i<j} (z_i-z_j)^{\lambda} \prod_{i} e^{-\frac{1}{4}|z_i|^2}.
\end{equation}
The background charge $e^{-i N \sqrt{\lambda} \phi(\infty)}$ ensures charge neutrality and leads to the Gaussian factor in \eqnref{eq:LJ_correlator}. As we restrict ourselves to the unit circle for our study of the CSM, this factor is a trivial constant and thus, the correlator coincides with the Laughlin--Jastrow factor \eqnref{eq:LJ_factor}.

The ground state can be used to transform the CSM Hamiltonian into the Laplace--Beltrami operator~\cite{AWATA1995347,lapointevinet,Bernevig_2008}
\begin{equation}
    \mathcal{H}_{\text{LB}}(1/\lambda)=\Psi_{\text{GS}}^{-1} ( \mathcal{H}_{\text{CSM}} - E_0) \Psi_{\text{GS}}\,,
\end{equation}
where
\begin{equation} \label{eq:Laplace_Beltrami}
  \mathcal{H}_{\text{LB}}(\alpha)
  = \sum_i \left(z_i  \partial_i \right)^2
  + \frac{1}{\alpha} \sum_{i<j} \frac{z_i + z_j}{z_i-z_j}
  \left(z_i  \partial_i - z_j  \partial_j \right)\,
\end{equation}
and
\begin{equation} \label{eq:energy_offset}
    E_0 = \frac{\lambda^2}{12} N(N-1)(N+1)
\end{equation}
is the ground state energy.
The eigenfunctions of \eqnref{eq:Laplace_Beltrami} are given by symmetric Jack polynomials, and describe the anyonic excitations of the CSM and fractional quantum Hall models~\cite{Bernevig_2008}.

Although the energy spectrum and the nature of the excitations of the CSM can be fully derived from \eqnref{eq:Laplace_Beltrami}, the operator formalism of the CSM offers a compact and elegant route to exposing its Hamiltonian and integrability structure. To this end, consider the annihilation operator 
\begin{align}\label{eq:CSM_GS_dest_op}
    D_i &= z_i \partial_i - \lambda \sum_{j\atop j\neq i} \frac{z_i}{z_i-z_j} + \frac{1}{2} \lambda (N-1)\nonumber\\
    &= z_i \partial_i - \frac{\lambda}{2}
    \sum_{j\atop j\neq i} \frac{z_i+z_j}{z_i-z_j}\,,
\end{align} 
for $i = 1,2,\dots,N$. As was realized in Refs.~\cite{lapointevinet,Bernevig_2008}, the CSM ground state \eqnref{eq:CSM_GS} lies in the kernel of these operators ($D_i\Psi_\mathrm{GS}=0$). Thus, the positive semidefinite bilinear $\sum_i D_i^{\dagger} D_i$ becomes a parent Hamiltonian for the ground state \eqnref{eq:CSM_GS} (details in Appendix~\ref{appA}):
\begin{align} \label{eq:parent_hamiltonian_CSM}
    \mathcal{H}_{\text{parent}} &= \sum_{i} \left(z_i\partial_i\right)^2-\lambda(\lambda-1) \sum_{i,j\atop i\neq j} \frac{z_iz_j}{(z_i-z_j)^2}\nonumber\\
    & \quad-\frac{1}{12}\lambda^2 N(N^2-1)\,.
\end{align}
This leads us to the standard result~\cite{10.1007/3-540-58453-6_3,lapointevinet} 
\begin{equation} \label{eq:hamiltonian_recipe}
    \mathcal{H}_{\rm CSM} = \sum_i D_i^{\dagger} D_i + E_0\,.
\end{equation}
This structure suggests that bilinear forms 
$$\mathcal{H}=\sum_i \Acsm_i^{\dagger} \Acsm_i + E_0$$
can be used to define parent Hamiltonians for states annihilated by the set $\{\Acsm_i\}$. While this construction guarantees that $\Psi_{\text{GS}}$ is a ground state, the uniqueness of this ground state is not guaranteed {\it a priori} and must be established separately in specific cases.

We note that systematic parent-Hamiltonian constructions for Jastrow and
Bijl--Jastrow wave functions have recently been developed in several continuum
settings, including confined one-dimensional systems, arbitrary-dimensional
pair-product Jastrow states, and graph-generalized Jastrow forms
\cite{Adolfo-2020,Beau-2021,Sasmal-2026}.
The construction developed below is complementary: rather than reconstructing
a Hamiltonian from a prescribed Jastrow amplitude alone, we use the BPZ
null-vector equations of the neutral RCFT correlator to generate
non-Abelian CSM-type annihilation operators.

Finally, this operator form is deeply connected to the integrability of the model. As demonstrated by Polychronakos~\cite{Polychronakos-1992}, one can construct a set of commuting conserved quantities using the exchange-operator formalism:
\begin{equation}
    \pi_i = z_i \partial_i - \frac{\lambda}{2}\sum_{j\atop j\neq i} \frac{z_i+z_j}{z_i-z_j} M_{ij}\,,
\end{equation}
where $M_{ij}$ is the coordinate permutation operator. The above operator can be used to construct ladder or step operators and reshape $\mathcal{H}_{\rm CSM}$ in the form of a harmonic-oscillator Hamiltonian~\cite{Brink_1992}.

\section{Construction of continuum models for \texorpdfstring{$\mathbf{c<1}$}{c<1} RCFTs}
\label{sec:csm_prescription}

We now detail the reverse-engineering framework used to derive our parent Hamiltonians. Our central assumption is that the trial state of interest can be represented as a conformal block of a primary field $\psi$ in a neutral RCFT with central charge $c < 1$ and a chiral boson CFT~\cite{Moore_1991}. We stress that, although the polynomial wave functions we target coincide with those of the corresponding two-dimensional FQH states, our construction places the particles on a circle and is therefore intrinsically one-dimensional: the natural physical setting is that of the Haldane--Shastry chain and its higher-spin generalizations~\cite{Greiter2011,PhysRevB.85.195149}, chiral spin liquids whose Gutzwiller-projected ground states are described by the very same Jastrow--Jack polynomials. We therefore use the FQH wave functions throughout merely as convenient analytic representatives of these one-dimensional chiral-spin-liquid states.

Prominent examples are provided by the Read--Rezayi $\mathbb{Z}_k$ series~\cite{Read-99prb8084}, which includes the Laughlin state ($k=1$) and the Moore--Read (Pfaffian) state ($k=2$). We define the FQH state as:
\begin{align}
\label{eq:1}
  \Psi(z)&=\langle \psi(z_1)\dots\psi(z_N)\rangle \Psi_{\text{LJ}}^\lambda(z)
           \nonumber\\[5pt]
    &\equiv \Psi_{\text{n}}(z)\Psi_{\text{LJ}}^\lambda(z)\,.
\end{align}
With each primary field $\psi$ of conformal dimension $h$ in a CFT with central charge $c$, we associate a primary state $\ket{\psi}=\psi(0)\ket{0}$. The corresponding Verma module is obtained by acting on this state with Virasoro generators $L_{-k}$ ($k >0$). At level $n$, the determinant of the Gram matrix $M_n$ is given by the Kac formula~\cite{ginsparg_1988,DiFrancesco_1997}:
\begin{equation}\label{eq:Kac}
    \det(M_n) \propto \prod_{p,q \geq 1 \atop pq \leq n} \left[ h-h_{p,q}(m)\right]^{P(n-pq)}\,,
\end{equation}
where $P(n)$ denotes the number of integer partitions of $n$, $c=1-\frac{6}{m(m+1)}$, and the conformal dimensions are:
\begin{equation}
h_{p,q}(m)=\frac{[(m+1)p-mq]^2-1}{4m(m+1)}\,.
\end{equation}
If $h=h_{p,q}(m)$, the existence of a null state at level $n = p \cdot q$ implies a linear combination of Virasoro generators such that:
\begin{equation}\label{eq:null_state_condition}
    \sum_{\substack{\scriptscriptstyle k \geq 1\\ \scriptscriptstyle n_1\geq \dots \geq n_k \geq 1\\ \scriptscriptstyle n_1+\dots+n_k=pq}} \alpha_{n_1,\dots,n_k} L_{-n_1}\dots L_{-n_k} \lvert \psi \rangle=0.
\end{equation}
The coefficients $\alpha$ are determined by requiring the state to be annihilated by all $L_n$ ($n>0$)~\cite{DiFrancesco_1997,ginsparg_1988}.

This null-state condition translates into a $pq$-order partial differential equation, the Belavin--Polyakov--Zamolodchikov (BPZ) equation, for the $N$-point correlation functions~\cite{1984NuPhB.241..333B,DiFrancesco_1997,AWATA1995347}:
\begin{equation}
    \A_i \langle\psi(z_1)\dots\psi(z_N)\rangle = 0\,,
\end{equation}
where the differential operator $\A_i$ is constructed from the Virasoro representations $\mathcal{L}_{-k}$. Specifically, 
$$\mathcal{L}_{-1} = \frac{\partial}{\partial z_i}\,,$$
and for $k \geq 2$:
\begin{equation}
  \mathcal{L}_{-k}
  =-\sum_{j\atop j\neq i}\left(\frac{(1-k)h}{(z_j-z_i)^k}
    +\frac{1}{(z_j-z_i)^{k-1}}\frac{\partial}{\partial z_j}\right)\,.
\end{equation}
 The differential operator $\A_i$ thus annihilates the neutral part of the FQH state:
\begin{equation}\label{eq:BPZ}
    0=\A_i\langle \psi(z_1)\dots\psi(z_N)\rangle = \A_i\Psi_{\text{n}}(z)\,.
\end{equation}

To map this to the full state $\Psi(z)$, we employ the product rule and the CSM annihilation operator $D_i$:
\begin{align}
    \Psi_{\text{LJ}}^\lambda z_i \frac{\partial \Psi_{\text{n}}}{\partial z_i}
    &= z_i \frac{\partial \Psi}{\partial z_i} - \Psi_{\text{n}} z_i \frac{\partial \Psi_{\text{LJ}}^\lambda}{\partial z_i} \nonumber\\
    &= \left(z_i \partial_i- \lambda \sum_{j \neq i} \frac{z_i}{z_i-z_j} \right)\Psi = D_i \Psi\,.
\end{align}
By replacing the momentum operators $z_i\partial_i$ in $\A_j$ with the annihilation operator of the CSM,
\begin{equation}
    D_i=z_i\partial_i-\lambda \sum_{j\atop j\neq i} \frac{z_i}{z_i-z_j}\,,
\end{equation}
we obtain an annihilation operator $\Acsm_j$ of the full state $\Psi(z)$.  By multiplying $\A_i$ with $z_j$'s, we can guarantee that only products $z_j\partial_j$ and no single derivative operators $\partial_j$ appear in the annihilation operator $\A_i$, which implies that we can replace all partial derivatives with the CSM annihilation operator. Then, by taking the positive semidefinite combination $\mathcal{H}=\sum_i\Acsm_i^\dag \Acsm_i$, we obtain a positive semidefinite Hamiltonian with $\Psi$ as a ground state, $\mathcal{H}\Psi=0$.

To summarize, we follow the following steps to obtain a parent Hamiltonian for a state given by the correlators of primary fields with a null vector at level $pq$:
\begin{enumerate}\label{eq:continuum_model_prescription}
    \item Construct the BPZ annihilation operator \eqnref{eq:BPZ} by finding the linear combination $\alpha_{n_1,\dots,n_k}$ annihilating the null state in \eqnref{eq:null_state_condition}.
    \item Multiply $\A_i$ with $z_j$'s, such that only products $z_j\partial_j$ appear in $\A_i$.
    \item Perform the substitution $z_j\partial_j\rightarrow D_j$ to obtain the annihilation operator $\Acsm_i$ of the full state $\Psi(z)$.
    \item Construct the parent Hamiltonian $\mathcal{H}=\sum_i \Acsm_i^\dag \Acsm_i$.
\end{enumerate}

While this construction guarantees $\Psi$ is a ground state of $\mathcal{H}$, the full energy spectrum and the potential degeneracy of the ground-state manifold remain open questions. Whether these Hamiltonians faithfully capture the anyon excitations of the corresponding chiral spin liquid system is a subject for future investigation.

For the $c<1$ unitary Virasoro minimal-model RCFTs considered here, the relevant Kac-table primary is degenerate and possesses a null vector at a sufficiently low level, from which the annihilation operator $\mathcal{A}_i$ is obtained. Even though these RCFTs have $c<1$, the essential requirement is not $c<1$ by itself, but the availability of a suitable null-vector annihilation equation. At the same time, for a generic primary in a CFT with $c \geq 1$, the Verma module is nondegenerate and no such null vector exists~\cite{ginsparg_1988,DiFrancesco_1997}, so our reverse-engineering procedure does not apply. This does not preclude a parent Hamiltonian by other means --- as exemplified by the Read--Rezayi series, whose neutral parafermion sector has $c \geq 1$ for $k \geq 4$ yet admits local $(k{+}1)$-body clustering Hamiltonians~\cite{Read-99prb8084}. Should one supply an annihilation operator $\A_i$ through other means such as clustering, steps 2-4 can still be followed to construct a parent Hamiltonian.

In Secs.~\ref{sec:pfaffian} and \ref{sec:rr}, we will construct the BPZ equations for the Moore--Read and the $k=3$ Read--Rezayi states to derive their corresponding annihilation operators $\Acsm_i$.

\section{Pfaffian continuum model}
\label{sec:pfaffian}
The Moore--Read (MR) state~\cite{Moore_1991} is a landmark trial wave function describing a non-Abelian fractional quantum Hall fluid at filling fraction $\nu = 1/2$. It is a leading candidate~\cite{greiter-91prl3205,greiter-92prb9586} for the observed $\nu = 5/2$ plateau, assuming the $2+1/2$ construction where the first two Landau levels are fully occupied and inert. Generalizing the Laughlin state by incorporating pairing correlations, the MR state captures the physics of a $p$-wave superconducting phase within the quantum Hall regime~\cite{greiter-92prb9586,Moore_1991}. Because of this particle pairing, the total particle number is always a multiple of two. Hence, its coordinate representation for even $N$ is given by:
\begin{equation} \label{eq:Moore--Read_state}
  \Psi_{\text{MR}}(z)
  = \text{Pf} \left( \frac{1}{z_i - z_j} \right) \prod_{i<j}(z_i - z_j)^{\lambda}
  \prod_{i} e^{-\frac{1}{4}|z_i|^2}\,,
\end{equation}
where $\lambda = 2$ corresponds to filling $\nu = 1/2$ in the second Landau level. The Pfaffian factor encodes $p$-wave pairing and---due to the Majorana zero modes associated with vortex pairs~\cite{Read_00}---gives rise to non-Abelian statistics. The MR state represents one of the simplest realizations of a non-Abelian topological phase in the continuum, providing a natural starting point for constructing a parent Hamiltonian in one dimension.

Variants of this state, such as the Read–Rezayi (RR) series, generalize the pairing to $k$-particle clusters using $\mathbb{Z}_k$ parafermion CFTs (formally the SU$(2)_k/$U(1) coset). These describe clustered quantum Hall states with increasingly complex non-Abelian structures~\cite{Read-99prb8084}. While the precise nature of the $\nu = 5/2$ ground state in experimental systems remains a subject of active debate---testing various candidates against thermal Hall conductance and tunneling data~\cite{greiter-91prl3205,greiter-92prb9586,Rezayi-00prl4685,nayak_08,Banerjee2018,lotric_2025}---the MR state remains the theoretical benchmark for non-Abelian order.

Similar to the Laughlin state \eqnref{eq:LJ_factor}, the MR state is intimately connected to CFT. Although the historical origin of the Laughlin state traces back to intuitive arguments for proposing a ground state {\it Ansatz} at the $\nu=\frac{1}{3}$ filling fraction, the Moore--Read state---with its coordinate representation, as given in \eqnref{eq:Moore--Read_state}---was originally motivated by CFT correlators. Specifically, its structure corresponds to the tensor product of the correlation function of an Ising CFT with a free boson, where the electron operator is represented by the product of an Ising Majorana fermion $\psi(z)$ and a free boson vertex operator:
\begin{align}
  \label{eq:MR_correlator}
     &\left \langle \prod_i \psi(z_i) :\!e^{i \sqrt{\lambda} \phi(z_i)}\!:e^{-i N \sqrt{\lambda} \phi(\infty)} \right \rangle \nonumber\\
    & \qquad = \text{Pf} \left( \frac{1}{z_i - z_j} \right) \prod_{i<j} (z_i - z_j)^\lambda \prod_{i} e^{-\frac{1}{4}|z_i|^2}\,. 
\end{align}

This construction not only motivates the analytic form of the wave function but also provides a framework for studying its quasihole excitations, which behave as Ising anyons obeying non-Abelian fusion rules~\cite{Moore_1991,Read_00,nayak_08}. These quasiholes can exhibit braiding statistics governed by the fusion algebra of the Ising CFT, which has been a central motivation for later theoretical and experimental studies of the $\nu = 5/2$ Hall plateau.

For the construction of a parent Hamiltonian in 1D, the Gaussian factor on the unit circle becomes a trivial constant and may be omitted. We can therefore write the Pfaffian state as 
\begin{equation}\label{eq:Pfaffian}
    \Psi_{\text{Pf}}(z)=\langle \psi(z_1)\dots \psi(z_N)\rangle \Psi_{\text{LJ}}^\lambda(z)\,.
\end{equation} 
The Majorana field $\psi(z)$ is the primary field with conformal dimension $h_{2,1}=1/2$ in the $c=1/2$ Ising CFT, corresponding to $m=3$ in the Kac formula. This implies the existence of a null state at level $2$, allowing us to apply the BPZ construction described in Sec.~\ref{sec:csm_prescription}. The null state is generated by
\begin{equation}
L_{-2}-\frac{3}{2(2h+1)}L_{-1}^2\, .
\end{equation}
This corresponds to the second-order differential equation on the $N$-point correlator:
\begin{align}
  \label{eq:BPZ_BP}
    &\biggl\{ \sum_{j\atop j\neq i}
      \bigg[\frac{h}{(z_i-z_j)^2} + \frac{1}{z_i-z_j}\frac{\partial}{\partial z_j}\bigg]
      \nonumber\\
    & \qquad - \frac{3}{2(2h+1)} \frac{\partial^2}{\partial z_i^2}\biggl\}\,
    \text{Pf} \left( \frac{1}{z_i - z_j} \right) = 0
\end{align}
for all $i\in\{1,2,\dots,N\}$. By invoking the translational invariance of the correlator, we can recast the resulting differential equation (see \appref{appB} for details) as
\begin{align}
  \label{eq:Pf_dest_op}
  \biggl\{ &\left(z_i\partial_i\right)^2 +\frac{1}{3} z_i \partial_i
             - \frac{4}{3}
             \sum_{j\atop j\neq i} \frac{z_i}{z_i-z_j} z_j\partial_j\nonumber\\
           & \qquad \quad - \frac{2}{3}
             \sum_{j\atop j\neq i} \frac{z_i^2}{(z_i-z_j)^2} \biggl\}
             \mathrm{Pf}\left(\frac{1}{z_i-z_j}\right) = 0\,.
\end{align}
Following the prescription in Sec.~\ref{sec:csm_prescription}, we replace the momentum operators $z_j\partial_j \to D_j$ to account for the Jastrow factor $\Psi_{\text{LJ}}^\lambda$. This yields the full annihilation operator for the Moore--Read state:
\begin{equation} \label{eq:BP_dest_op1}
    \Acsm_i^{\text{Pf}} = D_i^2 + \frac{1}{3} D_i - \frac{4}{3} \sum_{j\atop j\not=i} \frac{z_i}{z_i-z_j} D_j - \frac{2}{3} \sum_{j\atop j\not=i} \frac{z_i^2}{(z_i-z_j)^2}\,,
\end{equation}
where
\begin{equation}
    D_i = z_i \partial_i - \lambda \sum_{j\atop j\not=i} \frac{z_i}{z_i-z_j}\,.
\end{equation}
Expanding the $D_i$ operators and simplifying yields
\begin{align} \label{eq:expanded_BP_dest_op}
    \Acsm_i^{\text{Pf}} = &\left( z_i \partial_i \right)^2 - \left(-\lambda^2+ \frac{\lambda}{3}+\frac{2}{3} \right)
    \sum_{j\atop j\neq i} \frac{z_i^2}{z_{ij}^2} \nonumber\\
    &+ \left( \lambda^2+\frac{2}{3} \lambda \right)
    \sum_{j,k \atop i\neq j\neq k\neq i} \frac{z_i^2}{z_{ij}z_{ik}} + \frac{1}{3} z_i \partial_i
    \nonumber\\
    &-\sum_{j \atop j\neq i}
    \frac{z_i}{z_{ij}}\left( \frac{4}{3}  z_j \partial_j +2\lambda  z_i \partial_i\right)\,,
\end{align}
where $z_{kl}\equiv z_k - z_l$. Together with its adjoint $(\Acsm_i^{\text{Pf}})^{\dagger}$ [see~\eqnref{eq:BP_adjoint_op}], we propose the bilinear
$\sum_i (\Acsm_i^{\text{Pf}})^{\dagger} \Acsm_i^{\text{Pf}}$ as a parent Hamiltonian for the Pfaffian state \eqnref{eq:Pfaffian}.

\section{Fibonacci continuum model}\label{sec:rr}
Following the procedure established in Sec.~\ref{sec:pfaffian}, we now derive the annihilation operator for the $k=3$ Read--Rezayi (RR) state~\cite{Read-99prb8084}. Notably, $k=3$ is the only other theory in the $\mathbb{Z}_k$ parafermion series with a central charge strictly less than one ($c=4/5$) and still admitting a degenerate neutral primary field. This restricts our BPZ construction to $k \leq 3$, making the $k=3$ RR state the last model accessible within the present framework.

As a natural generalization of the Moore--Read state, the $k=3$ RR state describes a fractional quantum Hall phase of bosons at filling fraction $\nu= 3/2$, or fermions at $\nu = 3/5$. Its particle-hole conjugate at $\nu=2/5$ has long been discussed as a candidate description of the $\nu=2+2/5$ fractional quantum Hall plateau observed in the second Landau level~\cite{xia-04prl176809,Rezayi_2009}. The analytic form of this wave function is represented by the correlator:
\begin{equation}
    \Psi_{\text{RR3}}(z) = \langle \psi_1(z_1)\dots\psi_1(z_N)\rangle\Psi_{\text{LJ}}^{\lambda+2/3}(z)
\end{equation}
evaluated within the $c=4/5$, $\mathbb{Z}_3$ parafermion CFT~\cite{Read-99prb8084}, with $N$ being a multiple of $3$. The non-Abelian statistics of its quasiparticles are more intricate than those of the Moore–Read state: the fundamental excitations are Fibonacci anyons, whose fusion rules support universal topological quantum computation~\cite{nayak_08}.

The primary field $\psi_1$ possesses a conformal dimension $h_{1,3} = 2/3$. According to the Kac determinant, this dimension guarantees the existence of a null state at level $3$, generated by the Virasoro operator combination:
\begin{equation}
    L_{-3} - \frac{2}{h} L_{-2} L_{-1} + \frac{1}{h(h + 1)} L_{-1}^3\,.
\end{equation}
Translating this null-state condition into the coordinate representation yields the third-order BPZ differential equation for the neutral parafermion correlator $\mathcal{A}_i^{\rm RR3} \langle \psi_1(z_1) \dots \psi_1(z_N) \rangle =0$, with:
\begin{align}
  \label{eq:BPZ_parafermion}
  &\A_i^{\text{RR3}}=\frac{\partial^3}{\partial z_i^3} \nonumber\\
  &\quad - 2(h+1) \sum_{j\atop j\neq i}
  \left[\frac{h}{(z_i-z_j)^2} + \frac{1}{z_i-z_j}\frac{\partial}{\partial z_j} \right]
  \frac{\partial}{\partial z_i} \nonumber\\
  &\quad -h(h+1) \sum_{j\atop j\neq i} \left[ \frac{1}{(z_i-z_j)^2}
  \frac{\partial}{\partial z_j} + \frac{2h}{(z_i-z_j)^3}\right]\,.
\end{align}
Again, using translation invariance and replacing $z_j\partial_j$ with $D_j$ after substituting $h = 2/3$ (see \appref{appC}), we obtain the annihilation operator for the $k=3$ Read--Rezayi state
\begin{align}
  \label{eq:RR3_dest_op}
  \Acsm_i^{\text{RR}3}&= D_i^3 + \frac{1}{3} D_i^2 - \frac{2}{9} D_i\nonumber\\
  &- \frac{20}{9} \sum_{j\atop j\neq i} \frac{z_i^2}{(z_i-z_j)^2}D_i
    - \frac{10}{3} \sum_{j\atop j\neq i} \frac{z_i}{z_i-z_j}D_jD_i \nonumber\\
  &- \frac{10}{9} \sum_{j\atop j\neq i} \frac{z_i^2}{(z_i-z_j)^2}D_j
    - \frac{10}{9} \sum_{j\atop j\neq i} \frac{z_i}{z_i-z_j} D_j\nonumber\\
    & - \frac{40}{27} \sum_{j\atop j\neq i} \frac{z_i^3}{(z_i-z_j)^3}\,,
\end{align}
 where
\begin{equation}
    D_i = z_i \partial_i - \left(\lambda+\frac{2}{3}\right) \sum_{j\atop j\neq i}\frac{z_i}{z_i-z_j}
\end{equation}
is the CSM annihilation operator of $\Psi_{\text{LJ}}^{\lambda+2/3}$. In \appref{appC}, we present the expanded expression of \eqnref{eq:RR3_dest_op} and the form of its adjoint.

\section{Outlook}\label{sec:conclusion}
In this work, we have developed a general reverse-engineering approach to construct continuum parent Hamiltonians for one-dimensional chiral spin liquid trial states---equivalently, the polynomial parts of FQH states---derived from RCFTs with central charge $c<1$. Our framework demonstrates that the null-vector structure of an underlying RCFT can be systematically converted into many-body annihilation operators of Calogero--Sutherland type. We have explicitly demonstrated this construction for the Moore--Read and $k=3$ Read--Rezayi states, identifying the precise continuum Hamiltonians for which these non-Abelian states are exact zero modes. While our framework guarantees that the target wave function is an exact zero mode of the resulting Hamiltonian, it does not a priori guarantee uniqueness. Therefore, to assess the physical content of these models more fully, it is essential to investigate their energy spectra. Exact diagonalization of finite systems will therefore be important for determining the degeneracy of the ground-state manifold and verifying whether the anyonic quasiparticle excitations are accurately captured by these parent Hamiltonians. Being one-dimensional, these models are not expected to exhibit a genus-dependent topological ground-state degeneracy of the kind familiar from the two-dimensional FQH problem. The fractional and non-Abelian content of the corresponding chiral spin liquids instead manifests itself in one-dimensional diagnostics, such as the relative momentum spacing of the spinon excitations~\cite{PhysRevB.100.115107} and the structure of the entanglement spectrum of the ground state~\cite{Thomale_2010,Thomale_2011} and of its excitations. These, rather than a ground-state degeneracy, are the quantities against which the physical content of our parent Hamiltonians should ultimately be tested.

Crucially, our generalized annihilation operators are constructed directly from the standard CSM annihilation operators. This algebraic structure strongly suggests that the underlying commutation relations of the CSM might illuminate the properties of our extended models. A natural next step is to search for generalized Dunkl-like operators associated with our non-Abelian annihilation operators. Drawing inspiration from the exchange-operator formalism developed by Polychronakos~\cite{Polychronakos-1992}, modifying these operators to include coordinate permutations could provide the key to uncovering an infinite set of mutually commuting observables. The existence of such an integrability structure would allow for the generalization of the established operator solution formalism~\cite{10.1007/3-540-58453-6_3,lapointevinet,AWATA1995347} to analytically derive both the full energy spectrum and the Yangian monodromy matrices of these non-Abelian models.

Finally, extending the concept of ``freezing'' the continuous degrees of freedom at their classical static equilibrium points---which maps the continuum CSM to the \emph{lattice} Haldane--Shastry spin model~\cite{polychronakos93prl2329}, whose low-energy theory is in turn the $\mathrm{SU}(2)_1$ WZW model of free spinons interacting through their fractional statistics---presents a highly promising avenue for future research. In direct analogy, it is an open question whether continuum models of the type constructed here, upon freezing, reproduce the known higher-spin chiral-spin-liquid Hamiltonians~\cite{Greiter2011,PhysRevB.85.195149,PhysRevLett.99.097202,PhysRevB.80.104406,PhysRevLett.102.207203,PhysRevB.89.165125} or instead yield a genuinely new class of potentially solvable non-Abelian lattice models; resolving this is the central aim of future work.  Establishing such a link would help disentangle which features of the construction are genuinely continuum in nature and which survive in a lattice regularization.

More broadly, the present construction suggests that BPZ null-vector equations
may be viewed not only as constraints on conformal blocks, but as a practical
generating principle for many-body parent Hamiltonians. In this sense, it is
complementary to recent systematic constructions of parent Hamiltonians for
Jastrow-type wave functions, where the input is the pair-product structure of
the ground-state amplitude rather than the null-vector structure of an RCFT
correlator~\cite{Adolfo-2020,Beau-2021,Sasmal-2026}. A natural further question is whether this principle can be extended from the cases considered here to the broader hierarchy of $W_n$ algebras, thereby potentially extending our procedure to conformal field theories with Virasoro central charge $c>1$. This would clarify whether the Calogero--Sutherland paradigm is restricted to specific minimal-model settings or admits a broader extension to trial states associated with more general chiral algebras, perhaps at the price of more complicated, nonlocal, or matrix-valued annihilation operators. We hope the controlled route identified here serves as a starting point for exploring these richer spectral and algebraic structures.

\begin{acknowledgments} 
H.B. acknowledges the hospitality and financial support of the Chair for Theoretical Physics I, Julius-Maximilians-Universität Würzburg, where a large part of this work was carried out. The authors are indebted to Ronny Thomale and Martin Greiter for proposing the problem, formulating the idea, and initiating the project. A.F. acknowledges helpful discussions with F.D.M. Haldane and B.A. Bernevig.  H.B. and Y.I. acknowledge helpful discussions with S. Govindarajan, A. Balram, G. Mussardo and Z. Nussinov. This research was supported in part by grant NSF PHY-2309135 to the Kavli Institute for Theoretical Physics (KITP). The work of Y.I. was performed in part at the Aspen Center for Physics, which is supported by a grant from the Simons Foundation (1161654, Troyer). Y.I. acknowledges support from the Abdus Salam International Centre for Theoretical Physics through the Associates Programme, from the Simons Foundation through Grant No.~284558FY19, from IIT Madras through the Institute of Eminence program for establishing QuCenDiEM (Project No. SP22231244CPETWOQCDHOC), and the International Centre for Theoretical Sciences for participation in the Discussion Meeting --- Fractionalized Quantum Matter (code: ICTS/DMFQM2025/07). All authors acknowledge the Deutsche Forschungsgemeinschaft (DFG, German Research Foundation) through Project-ID 258499086-SFB 1170 and the W\"urzburg-Dresden Cluster of Excellence on Complexity and Topology in Quantum Matter – ct.qmat Project-ID 390858490-EXC 2147. 
\end{acknowledgments}


\appendix
\section{CSM calculations} \label{appA}
We consider particles on the unit circle, parameterized by $z_k=e^{ix_k}$, and define the standard inner product on the unit circle,
\[
\langle \eta_1|\eta_2\rangle
=
\left(\frac{1}{2\pi i}\right)^N
\int \prod_j \frac{dz_j}{z_j}\,
\eta_1^*(\{z_j\})\,\eta_2(\{z_j\})\,.
\]
With respect to this inner product, the Euler operator $z_k\partial_k$ is self-adjoint and corresponds to the one-dimensional momentum operator $-i\partial_{x_k}$. Conjugation acts on the coordinates as complex conjugation, which we denote by $z\to \bar{z}$. Consequently, the adjoint of the annihilation operator $D_i$ defined in \eqnref{eq:CSM_GS_dest_op} differs only by a sign in the second term:
\begin{equation} \label{eq:CSM_GS_dest_adjoint}
\begin{aligned}
    D_i^{\dagger} &=  z_i \partial_i - \lambda \sum_{j \atop j\neq i} \frac{\bar{z_i}}{\bar{z_i}-\bar{z_j}} + \frac{1}{2} \lambda (N-1)\\
    &= z_i \partial_i - \lambda \sum_{j\atop j \neq i} \frac{z_j}{z_j-z_i} + \frac{1}{2} \lambda (N-1)\\&= z_i \partial_i + \frac{\lambda}{2} \sum_{j \atop j\neq i} \frac{z_i+z_j}{z_i-z_j}\,.
\end{aligned}
\end{equation}
The bilinear formed from the annihilation operator and its adjoint yields the CSM Hamiltonian \eqnref{eq:CSM_hamiltonian} up to an additive constant. Expanding the product, we obtain:
\begin{widetext}
\begin{equation}
\begin{aligned}
    \sum_{i} D^{\dagger}_i D_i &= \sum_{i} \left(z_i\partial_i+\frac{\lambda}{2}\sum_{k \atop k\neq i}\frac{z_i+z_k}{z_i-z_k}\right) \left(z_i\partial_i-\frac{\lambda}{2}\sum_{j \atop j\neq i}\frac{z_i+z_j}{z_i-z_j}\right)\\
    &= \sum_{i} \Biggl[ (z_i\partial_i)^2 - \lambda(\lambda-1)\sum_{j \atop j\neq i} \frac{z_iz_j}{(z_i-z_j)^2} - \frac{\lambda^2(N-1)^2}{4} - \lambda^2\sum_{j,k\atop i\neq j\neq k\neq i}\frac{z_iz_j}{(z_i-z_k)(z_i-z_j)} \Biggr]\\
    &= \sum_{i} (z_i\partial_i)^2 - \lambda(\lambda-1)\sum_{i,j\atop i \neq j} \frac{z_iz_j}{(z_i-z_j)^2} - \frac{1}{12}\lambda^2 N(N^2-1)\,,
\end{aligned}
\end{equation}
\end{widetext}
which recovers the result \eqnref{eq:parent_hamiltonian_CSM} in the main text.

\section{Annihilation operator of the Pfaffian} \label{appB}
In this section, we provide a derivation of the Pfaffian annihilation operator $\Acsm_i^{\text{Pf}}$ given in \eqnref{eq:BP_dest_op1}. We adopt the shorthand notation $\partial_i \equiv \frac{\partial}{\partial z_i}$ and $z_{ij} \equiv z_i - z_j$.

We begin by invoking the translational invariance of the Pfaffian correlator:
\begin{align} \label{eq:Pf_translation}
    0 &= \biggl\{ \partial_i + \sum_{j\atop j \neq i} \partial_j \biggl\} \mathrm{Pf}\left( \frac{1}{z_i-z_j}\right)\nonumber\\
    &= \biggl\{ \partial_i +\sum_{j\atop j \neq i}  \frac{z_i-z_j}{z_i-z_j} \partial_j \biggl\} \mathrm{Pf}\left( \frac{1}{z_i-z_j}\right)\nonumber\\
    &= \biggl\{ \partial_i + \sum_{j \atop j\neq i} \frac{z_i}{z_{ij}} \partial_j -  \sum_{j \atop j\neq i} \frac{z_j}{z_{ij}}\partial_j\biggl\} \mathrm{Pf}\left( \frac{1}{z_i-z_j}\right)\,.
\end{align}
Multiplying the second-order differential equation \eqnref{eq:BPZ_BP} by $- 4 z_i^2 / 3$ and adding $4 z_i / 3$ times \eqnref{eq:Pf_translation}, we arrive at \eqnref{eq:Pf_dest_op}:
\begin{align}
  \biggl\{ &\left(z_i\partial_i\right)^2 +\frac{1}{3} z_i \partial_i
             - \frac{4}{3}
             \sum_{j\atop j\neq i} \frac{z_i}{z_i-z_j} z_j\partial_j\nonumber\\
           & \qquad \quad - \frac{2}{3}
             \sum_{j\atop j\neq i} \frac{z_i^2}{(z_i-z_j)^2} \biggl\}
             \mathrm{Pf}\left(\frac{1}{z_i-z_j}\right) = 0\,.
\end{align}

After simplification, the adjoint of \eqnref{eq:BP_dest_op1} is given by:
\begin{align} \label{eq:BP_adjoint_op}
    (\Acsm_i^{\text{Pf}})^{\dagger} &= \left( z_i \partial_i \right)^2
    - \left( -\lambda^2+\frac{\lambda}{3}+\frac{2}{3} \right) \sum_{j\atop j\neq i}  \frac{z_j^2}{z_{ji}^2}\nonumber\\
    &\quad +\left(\lambda^2+\frac{2}{3}\lambda\right) \sum_{j,k\atop i\neq j\neq k \neq i}  \frac{z_j z_k}{z_{ji} z_{ki}} 
    + \frac{1}{3}
    z_i \partial_i \nonumber\\
    &\quad + 2\left( \frac{2}{3} - \lambda \right) \sum_{j\atop j\neq i} \frac{z_iz_j}{z_{ji}^2}
    - \frac{4}{3} \sum_{j\atop j\neq i} \frac{z_j}{z_{ji}} z_j \partial_j  \nonumber\\
    &\quad- 2\lambda \sum_{j\atop j\neq i}  \frac{z_j}{z_{ji}}  z_i \partial_i\,.
\end{align}

\section{Annihilation operator of the \texorpdfstring{$\mathbf{k=3}$}{k=3} Read--Rezayi state}\label{appC}
To derive \eqnref{eq:RR3_dest_op} from \eqnref{eq:BPZ_parafermion}, we need to bring the annihilation operator $\A_i$ into a form, such that only terms $z_j\partial_j$ appear in it and we can perform the substitution $z_j\partial_j \to D_j$. Throughout this appendix, all operators are understood to act on a translationally invariant wave function. To this end, we multiply \eqnref{eq:BPZ_parafermion} by $z_i^3$:
\begin{align}\label{eq:A_RR3}
    \A_i^{\text{RR3}}&=z_i^3 \partial_i^3 - 2h(h+1) \sum_{j\atop j\neq i}
   \frac{z_i^2}{(z_i-z_j)^2} z_i \partial_i\nonumber\\
    &\quad- 2(h+1) \sum_{j\atop j\neq i} \frac{z_i}{z_i-z_j} (z_i \partial_j) (z_i \partial_i) \nonumber\\
    &\quad-h(h+1) \sum_{j\atop j\neq i} \left[\frac{z_i^2}{(z_i-z_j)^2}
    z_i \partial_j +\frac{2hz_i^3}{(z_i-z_j)^3}\right]\,.
\end{align}
To express mixed-index operators $z_i\partial_j$ in terms of the
momentum operators $z_i\partial_i$, we establish a set of intermediate identities. First, we rewrite the third-order and second-order Euler-like derivatives:
\begin{align} \label{eq:euler_identities}
z_i^2 \partial_i^2 &= (z_i \partial_i)^2 - z_i \partial_i \,, \nonumber \\
z_i^3 \partial_i^3 &= (z_i \partial_i)^3 - 3 (z_i \partial_i)^2 + 2 z_i \partial_i \, .
\end{align}

Next, we simplify the mixed-derivative term in line two of \eqnref{eq:A_RR3}:
\begin{align} \label{eq:intermediate_result_1}
    \sum_{j \atop j \neq i} & \frac{z_i}{z_i-z_j} (z_i \partial_j) (z_i \partial_i) = \sum_{j \atop j \neq i} \frac{z_i^2}{z_i-z_j} \partial_j (z_i \partial_i) \nonumber \\
    & = \sum_{j \atop j \neq i} \frac{z_i^2 - z_i z_j + z_i z_j}{z_i-z_j} \partial_j (z_i \partial_i) \nonumber \\
    & = \sum_{j \atop j \neq i} (z_i \partial_j) (z_i \partial_i) + \sum_{j \atop j \neq i} \frac{z_i}{z_i-z_j} (z_j \partial_j) (z_i \partial_i) \nonumber \\
    & = z_i^2 \partial_i \sum_{j \atop j \neq i} \partial_j + \sum_{j \atop j \neq i} \frac{z_i}{z_i-z_j} (z_j \partial_j) (z_i \partial_i) \nonumber \\
    & = -z_i^2\partial_i^2 + \sum_{j \atop j \neq i} \frac{z_i}{z_i-z_j} (z_j \partial_j) (z_i \partial_i) \nonumber \\
    & = -(z_i \partial_i)^2 + z_i \partial_i + \sum_{j \atop j \neq i} \frac{z_i}{z_i-z_j} (z_j \partial_j) (z_i \partial_i) \,.
\end{align}
In the penultimate step of \eqnref{eq:intermediate_result_1}, we utilize the translational invariance of the wave function.

Finally, the mixed derivative in line three of \eqnref{eq:BPZ_parafermion} simplifies to:
\begin{align} \label{eq:intermediate_result_2}
\sum_{j \atop j\neq i} &\frac{z_i^2}{(z_i - z_j)^2} z_i \partial_j = \sum_{j \atop j \neq i} \frac{z_i^3}{(z_i-z_j)^2} \partial_j \nonumber \\
& = \sum_{j \atop j \neq i} \frac{z_i^3 - z_i^2 z_j + z_i^2 z_j}{(z_i-z_j)^2} \partial_j \nonumber \\
& = \sum_{j \atop j \neq i} \frac{z_i^2}{(z_i-z_j)} \partial_j + \sum_{j \atop j \neq i} \frac{z_i^2}{(z_i-z_j)^2} z_j \partial_j  \nonumber \\
& = \sum_{j \atop j \neq i} \frac{z_i^2 - z_i z_j + z_i z_j}{(z_i-z_j)} \partial_j + \sum_{j \atop j \neq i} \frac{z_i^2}{(z_i-z_j)^2} z_j \partial_j  \nonumber \\
& = \sum_{j \atop j \neq i} z_i \partial_j + \sum_{j \atop j \neq i} \frac{z_i}{(z_i-z_j)} z_j \partial_j + \sum_{j \atop j \neq i} \frac{z_i^2}{(z_i-z_j)^2} z_j \partial_j  \nonumber \\
& = -z_i \partial_i + \sum_{j \atop j \neq i} \frac{z_i}{(z_i-z_j)} z_j \partial_j + \sum_{j \atop j \neq i} \frac{z_i^2}{(z_i-z_j)^2} z_j \partial_j \, .
\end{align}

Substituting Eqs.~\eqref{eq:euler_identities}--\eqref{eq:intermediate_result_2} into \eqnref{eq:A_RR3} and setting $h=2/3$, we obtain
\begin{align}
    \A_i^{\text{RR3}}&=(z_i \partial_i)^3 +\frac{1}{3} (z_i \partial_i)^2 -\frac{2}{9} (z_i \partial_i ) \nonumber\\
    &\quad- \frac{20}{9} \sum_{j \atop j \neq i} \frac{z_i^2}{(z_i-z_j)^2}\, (z_i \partial_i) \nonumber\\
    &\quad -\frac{10}{3}\left( \sum_{j \atop j \neq i} \frac{z_i}{z_i-z_j}\, (z_j \partial_j) (z_i \partial_i) \right) \nonumber \\
    &\quad - \frac{10}{9} \left( \sum_{j \atop j \neq i} \frac{z_i^2}{(z_i-z_j)^2}\, z_j \partial_j + \sum_{j \atop j \neq i} \frac{z_i}{z_i-z_j}\, z_j \partial_j\right) \nonumber \\
    & \quad - \frac{40}{27}\sum_{j \atop j \neq i} \frac{z_i^3}{(z_i-z_j)^3}\,.
\end{align}

Substituting $z_i\partial_i \to D_i$ in the above expression then yields \eqnref{eq:RR3_dest_op} with 
\begin{equation}
    D_i=z_i\partial_i-\left(\lambda+\frac{2}{3}\right)\sum_{j\atop j\neq i} \frac{z_i}{z_i-z_j}\,.
\end{equation}

For completeness, we record the fully expanded form of the operator. Defining
\[
A=\lambda+\frac{2}{3},
\]
we obtain:
\begin{widetext}
    \begin{align} \label{eq:expanded_RR3_dest_op}
        \Acsm_i^{\text{RR3}} =& (z_i \partial_{i})^3 -\frac{2}{9} z_i \partial_{i} 
        + \frac{1}{3} (z_i \partial_{i})^2 
        - \frac{(40 - 66A - 9A^2 + 27A^3)}{27} \sum_{j\atop j\neq i}\frac{z_i^3}{z_{ij}^3} 
        + \frac{(-20 - 3A + 27A^2)}{9} \sum_{j\atop j\neq i} \frac{z_i^2}{z_{ij}^2} (z_i \partial_{i}) \nonumber\\
        & + \frac{10(3A-1)}{9} \sum_{j\atop j\neq i} \frac{z_i^2}{z_{ij}^2} (z_j \partial_{j}) 
        - \frac{A}{3} \sum_{j\atop j\neq i} \frac{z_i}{z_{ij}} (z_i \partial_{i})
        - 3A \sum_{j\atop j\neq i} \frac{z_i}{z_{ij}} (z_i \partial_{i})^2
        - \frac{10}{9} \sum_{j\atop j\neq i} \frac{z_i}{z_{ij}} (z_j \partial_{j}) \nonumber\\
        & - \frac{10}{3} \sum_{j\atop j\neq i} \frac{z_i}{z_{ij}} (z_i \partial_{i})(z_j \partial_{j}) 
        - \frac{A (9A + 9A^2 - 10)}{3} \sum_{j,k\atop i\neq j \neq k \neq i} \frac{z_i^3}{z_{ij} z_{ik}^2} + \frac{A (5 + 9A)}{3} \sum_{j,k\atop i\neq j \neq k \neq i} \frac{z_i^2}{z_{ij} z_{ik}} (z_i \partial_{i}) \nonumber\\
        & + \frac{10}{3}A \sum_{j,k\atop i\neq j \neq k \neq i} \frac{z_i^2}{z_{ij} z_{ik}} (z_j \partial_{j}) 
        - \frac{A^2 (5 + 3A)}{3} \sum_{\substack{\scriptscriptstyle j,k,l \\\scriptscriptstyle i\neq j \neq k \neq i \\\scriptscriptstyle l\neq i,j,k}} \frac{z_i^3}{z_{ij} z_{ik} z_{il}}\,.
    \end{align}
The corresponding adjoint operator takes the form
\begin{align}
    (&\Acsm_i^{\text{RR3}})^{\dagger} = -\frac{1}{27} (5 + 3A)\Big[8 - 18A(-1 + N) + 9A^2(-1 + N)^2\Big](-1 + N) \nonumber \\
    &+ \frac{1}{9} \Big[18 + 27A^2(-1 + N)^2 - 20N + 3A(12 - 17N + 5N^2)\Big] (z_i \partial_i) + \frac{1}{3} \Big[1 - 9A(-1 + N)\Big] (z_i \partial_i)^2 + (z_i \partial_i)^3 \nonumber \\
    &+ \frac{10}{9} \Big[-2 + 3A(-1 + N) \Big] \sum_{j \atop j \neq i} z_j \partial_j - \frac{10}{3} \sum_{j \atop j \neq i} (z_i \partial_i) (z_j \partial_j) + \frac{(-80 + 294A - 171A^2 + 27A^3)}{27} \sum_{j \atop j \neq i} \frac{z_i^3}{z_{ij}^3} \nonumber \\
    &+ \frac{\big[20 + 30A(-7 + N) - 27A^3(-1 + N) + 9A^2(7 + 3N)\big]}{9} \sum_{j \atop j \neq i} \frac{z_i^2}{z_{ij}^2} + \frac{(10 - 57A + 27A^2)}{9} \sum_{j \atop j \neq i} \frac{z_i^2}{ z_{ij}^2} z_i \partial_i \nonumber \\
    &+ \frac{10(-4 + 3A)}{9 } \sum_{j \atop j \neq i} \frac{ z_i^2}{z_{ij}^2}  z_j \partial_j+ \frac{\big[20 + 27A^3(-1 + N)^2 - 30A(-7 + 3N) + 9A^2(17 - 22N + 5N^2)\big] }{9 } \sum_{j \atop j \neq i} \frac{z_i}{z_{ij}} \nonumber \\
    &+ \frac{\big[10 + A(123 - 30N) - 54A^2(-1 + N) \big] }{9 } \sum_{j \atop j \neq i} \frac{z_i}{z_{ij}} z_i \partial_i + 3A \sum_{j \atop j \neq i} \frac{z_i}{z_{ij}} (z_i \partial_i)^2 - \frac{10(-2 + AN)}{3} \sum_{j \atop j \neq i} \frac{z_i}{z_{ij}} z_j \partial_j  \nonumber \\
    &+ \frac{10}{3} \sum_{j \atop j \neq i} \frac{z_i}{z_{ij}} (z_i \partial_i) (z_j \partial_j) + \frac{A(-10 - 9A + 9A^2)}{3} \sum_{j,k \atop i \neq j \neq k \neq i} \frac{z_i^3}{z_{ij}z_{ik}^2} - \frac{10A}{3} \sum_{j,k\atop i\neq j \neq k \neq i} \frac{z_i}{z_{ik}} z_j \partial_j \nonumber \\
    &+ \frac{A \big[ 20 - 15A(-3 + N) - 9A^2(-1 + N)\big]}{3} \sum_{j,k\atop i\neq j \neq k \neq i} \frac{z_i^2}{z_{ij}z_{ik}} + \frac{A(5 + 9A)}{3} \sum_{j,k\atop i\neq j \neq k \neq i} \frac{z_i^2}{z_{ij}z_{ik}} z_i \partial_i \nonumber \\
    &+ \frac{10A}{3} \sum_{j,k\atop i\neq j \neq k \neq i} \frac{z_i^2}{z_{ij}z_{ik}} z_j \partial_j + \frac{A^2(5 + 3A)}{3} \sum_{\substack{\scriptscriptstyle j,k,l \\\scriptscriptstyle i\neq j \neq k \neq i \\\scriptscriptstyle l\neq i,j,k}} \frac{z_i^3}{z_{ij}z_{ik} z_{il}}\,.
\end{align}
The continuum parent Hamiltonian for the bosonic Fibonacci $k=3$ Read--Rezayi state is therefore the positive semidefinite bilinear
\[
\sum_i (\Acsm_i^{\text{RR3}})^{\dagger}\Acsm_i^{\text{RR3}}.
\]
\end{widetext}


\begin{thebibliography}{56}%
\makeatletter
\providecommand \@ifxundefined [1]{%
 \@ifx{#1\undefined}
}%
\providecommand \@ifnum [1]{%
 \ifnum #1\expandafter \@firstoftwo
 \else \expandafter \@secondoftwo
 \fi
}%
\providecommand \@ifx [1]{%
 \ifx #1\expandafter \@firstoftwo
 \else \expandafter \@secondoftwo
 \fi
}%
\providecommand \natexlab [1]{#1}%
\providecommand \enquote  [1]{``#1''}%
\providecommand \bibnamefont  [1]{#1}%
\providecommand \bibfnamefont [1]{#1}%
\providecommand \citenamefont [1]{#1}%
\providecommand \href@noop [0]{\@secondoftwo}%
\providecommand \href [0]{\begingroup \@sanitize@url \@href}%
\providecommand \@href[1]{\@@startlink{#1}\@@href}%
\providecommand \@@href[1]{\endgroup#1\@@endlink}%
\providecommand \@sanitize@url [0]{\catcode `\\12\catcode `\$12\catcode `\&12\catcode `\#12\catcode `\^12\catcode `\_12\catcode `\%12\relax}%
\providecommand \@@startlink[1]{}%
\providecommand \@@endlink[0]{}%
\providecommand \url  [0]{\begingroup\@sanitize@url \@url }%
\providecommand \@url [1]{\endgroup\@href {#1}{\urlprefix }}%
\providecommand \urlprefix  [0]{URL }%
\providecommand \Eprint [0]{\href }%
\providecommand \doibase [0]{https://doi.org/}%
\providecommand \selectlanguage [0]{\@gobble}%
\providecommand \bibinfo  [0]{\@secondoftwo}%
\providecommand \bibfield  [0]{\@secondoftwo}%
\providecommand \translation [1]{[#1]}%
\providecommand \BibitemOpen [0]{}%
\providecommand \bibitemStop [0]{}%
\providecommand \bibitemNoStop [0]{.\EOS\space}%
\providecommand \EOS [0]{\spacefactor3000\relax}%
\providecommand \BibitemShut  [1]{\csname bibitem#1\endcsname}%
\let\auto@bib@innerbib\@empty
\bibitem [{\citenamefont {Wen}(2017)}]{RevModPhys.89.041004}%
  \BibitemOpen
  \bibfield  {author} {\bibinfo {author} {\bibfnamefont {X.-G.}\ \bibnamefont {Wen}},\ }\bibfield  {title} {\bibinfo {title} {{Colloquium: Zoo of quantum-topological phases of matter}},\ }\href {https://doi.org/10.1103/RevModPhys.89.041004} {\bibfield  {journal} {\bibinfo  {journal} {Rev. Mod. Phys.}\ }\textbf {\bibinfo {volume} {89}},\ \bibinfo {pages} {041004} (\bibinfo {year} {2017})}\BibitemShut {NoStop}%
\bibitem [{\citenamefont {Greiter}\ and\ \citenamefont {Wilczek}(2024)}]{greiter-24arcmp131}%
  \BibitemOpen
  \bibfield  {author} {\bibinfo {author} {\bibfnamefont {M.}~\bibnamefont {Greiter}}\ and\ \bibinfo {author} {\bibfnamefont {F.}~\bibnamefont {Wilczek}},\ }\bibfield  {title} {\bibinfo {title} {{Fractional Statistics}},\ }\href {https://doi.org/https://doi.org/10.1146/annurev-conmatphys-040423-014045} {\bibfield  {journal} {\bibinfo  {journal} {Annu. Rev. Condens. Matter Phys.}\ }\textbf {\bibinfo {volume} {15}},\ \bibinfo {pages} {131} (\bibinfo {year} {2024})}\BibitemShut {NoStop}%
\bibitem [{\citenamefont {Moore}\ and\ \citenamefont {Read}(1991)}]{Moore_1991}%
  \BibitemOpen
  \bibfield  {author} {\bibinfo {author} {\bibfnamefont {G.}~\bibnamefont {Moore}}\ and\ \bibinfo {author} {\bibfnamefont {N.}~\bibnamefont {Read}},\ }\bibfield  {title} {\bibinfo {title} {{Nonabelions in the fractional quantum hall effect}},\ }\href {https://doi.org/https://doi.org/10.1016/0550-3213(91)90407-O} {\bibfield  {journal} {\bibinfo  {journal} {Nucl. Phys. B}\ }\textbf {\bibinfo {volume} {360}},\ \bibinfo {pages} {362} (\bibinfo {year} {1991})}\BibitemShut {NoStop}%
\bibitem [{\citenamefont {Nayak}\ \emph {et~al.}(2008)\citenamefont {Nayak}, \citenamefont {Simon}, \citenamefont {Stern}, \citenamefont {Freedman},\ and\ \citenamefont {Das~Sarma}}]{nayak_08}%
  \BibitemOpen
  \bibfield  {author} {\bibinfo {author} {\bibfnamefont {C.}~\bibnamefont {Nayak}}, \bibinfo {author} {\bibfnamefont {S.~H.}\ \bibnamefont {Simon}}, \bibinfo {author} {\bibfnamefont {A.}~\bibnamefont {Stern}}, \bibinfo {author} {\bibfnamefont {M.}~\bibnamefont {Freedman}},\ and\ \bibinfo {author} {\bibfnamefont {S.}~\bibnamefont {Das~Sarma}},\ }\bibfield  {title} {\bibinfo {title} {{Non-Abelian anyons and topological quantum computation}},\ }\href {https://doi.org/10.1103/RevModPhys.80.1083} {\bibfield  {journal} {\bibinfo  {journal} {Rev. Mod. Phys.}\ }\textbf {\bibinfo {volume} {80}},\ \bibinfo {pages} {1083} (\bibinfo {year} {2008})}\BibitemShut {NoStop}%
\bibitem [{\citenamefont {Khare}(1998)}]{khare2005fractional}%
  \BibitemOpen
  \bibfield  {author} {\bibinfo {author} {\bibfnamefont {A.}~\bibnamefont {Khare}},\ }\href {https://doi.org/10.1142/5752} {\emph {\bibinfo {title} {Fractional Statistics and Quantum Theory}}}\ (\bibinfo  {publisher} {World Scientific},\ \bibinfo {year} {1998})\ \Eprint {https://arxiv.org/abs/https://www.worldscientific.com/doi/pdf/10.1142/2988} {https://www.worldscientific.com/doi/pdf/10.1142/2988} \BibitemShut {NoStop}%
\bibitem [{\citenamefont {Gromov}\ and\ \citenamefont {Radzihovsky}(2024)}]{RevModPhys.96.011001}%
  \BibitemOpen
  \bibfield  {author} {\bibinfo {author} {\bibfnamefont {A.}~\bibnamefont {Gromov}}\ and\ \bibinfo {author} {\bibfnamefont {L.}~\bibnamefont {Radzihovsky}},\ }\bibfield  {title} {\bibinfo {title} {{Colloquium: Fracton matter}},\ }\href {https://doi.org/10.1103/RevModPhys.96.011001} {\bibfield  {journal} {\bibinfo  {journal} {Rev. Mod. Phys.}\ }\textbf {\bibinfo {volume} {96}},\ \bibinfo {pages} {011001} (\bibinfo {year} {2024})}\BibitemShut {NoStop}%
\bibitem [{\citenamefont {Haldane}(1991)}]{PhysRevLett.67.937}%
  \BibitemOpen
  \bibfield  {author} {\bibinfo {author} {\bibfnamefont {F.~D.~M.}\ \bibnamefont {Haldane}},\ }\bibfield  {title} {\bibinfo {title} {{``Fractional statistics'' in arbitrary dimensions: A generalization of the Pauli principle}},\ }\href {https://doi.org/10.1103/PhysRevLett.67.937} {\bibfield  {journal} {\bibinfo  {journal} {Phys. Rev. Lett.}\ }\textbf {\bibinfo {volume} {67}},\ \bibinfo {pages} {937} (\bibinfo {year} {1991})}\BibitemShut {NoStop}%
\bibitem [{\citenamefont {Ha}(1994)}]{PhysRevLett.73.1574}%
  \BibitemOpen
  \bibfield  {author} {\bibinfo {author} {\bibfnamefont {Z.~N.~C.}\ \bibnamefont {Ha}},\ }\bibfield  {title} {\bibinfo {title} {{Exact Dynamical Correlation Functions of Calogero-Sutherland Model and One-Dimensional Fractional Statistics}},\ }\href {https://doi.org/10.1103/PhysRevLett.73.1574} {\bibfield  {journal} {\bibinfo  {journal} {Phys. Rev. Lett.}\ }\textbf {\bibinfo {volume} {73}},\ \bibinfo {pages} {1574} (\bibinfo {year} {1994})}\BibitemShut {NoStop}%
\bibitem [{\citenamefont {Greiter}(2009)}]{PhysRevB.79.064409}%
  \BibitemOpen
  \bibfield  {author} {\bibinfo {author} {\bibfnamefont {M.}~\bibnamefont {Greiter}},\ }\bibfield  {title} {\bibinfo {title} {{Statistical phases and momentum spacings for one-dimensional anyons}},\ }\href {https://doi.org/10.1103/PhysRevB.79.064409} {\bibfield  {journal} {\bibinfo  {journal} {Phys. Rev. B}\ }\textbf {\bibinfo {volume} {79}},\ \bibinfo {pages} {064409} (\bibinfo {year} {2009})}\BibitemShut {NoStop}%
\bibitem [{\citenamefont {Greiter}(2011)}]{Greiter2011}%
  \BibitemOpen
  \bibfield  {author} {\bibinfo {author} {\bibfnamefont {M.}~\bibnamefont {Greiter}},\ }\href {https://doi.org/10.1007/978-3-642-24384-4} {\emph {\bibinfo {title} {Mapping of Parent Hamiltonians: From Abelian and non-Abelian Quantum Hall States to Exact Models of Critical Spin Chains}}}\ (\bibinfo  {publisher} {Springer Berlin Heidelberg},\ \bibinfo {address} {Heidelberg},\ \bibinfo {year} {2011})\BibitemShut {NoStop}%
\bibitem [{\citenamefont {Greiter}\ \emph {et~al.}(2019)\citenamefont {Greiter}, \citenamefont {Haldane},\ and\ \citenamefont {Thomale}}]{PhysRevB.100.115107}%
  \BibitemOpen
  \bibfield  {author} {\bibinfo {author} {\bibfnamefont {M.}~\bibnamefont {Greiter}}, \bibinfo {author} {\bibfnamefont {F.~D.~M.}\ \bibnamefont {Haldane}},\ and\ \bibinfo {author} {\bibfnamefont {R.}~\bibnamefont {Thomale}},\ }\bibfield  {title} {\bibinfo {title} {{Non-Abelian statistics in one dimension: Topological momentum spacings and SU(2) level-$k$ fusion rules}},\ }\href {https://doi.org/10.1103/PhysRevB.100.115107} {\bibfield  {journal} {\bibinfo  {journal} {Phys. Rev. B}\ }\textbf {\bibinfo {volume} {100}},\ \bibinfo {pages} {115107} (\bibinfo {year} {2019})}\BibitemShut {NoStop}%
\bibitem [{\citenamefont {Haldane}(1983)}]{haldane83prl605}%
  \BibitemOpen
  \bibfield  {author} {\bibinfo {author} {\bibfnamefont {F.~D.~M.}\ \bibnamefont {Haldane}},\ }\bibfield  {title} {\bibinfo {title} {{Fractional Quantization of the Hall Effect: A Hierarchy of Incompressible Quantum Fluid States}},\ }\href {https://doi.org/10.1103/PhysRevLett.51.605} {\bibfield  {journal} {\bibinfo  {journal} {Phys. Rev. Lett.}\ }\textbf {\bibinfo {volume} {51}},\ \bibinfo {pages} {605} (\bibinfo {year} {1983})}\BibitemShut {NoStop}%
\bibitem [{\citenamefont {Thomale}\ \emph {et~al.}(2011)\citenamefont {Thomale}, \citenamefont {Estienne}, \citenamefont {Regnault},\ and\ \citenamefont {Bernevig}}]{PhysRevB.84.045127}%
  \BibitemOpen
  \bibfield  {author} {\bibinfo {author} {\bibfnamefont {R.}~\bibnamefont {Thomale}}, \bibinfo {author} {\bibfnamefont {B.}~\bibnamefont {Estienne}}, \bibinfo {author} {\bibfnamefont {N.}~\bibnamefont {Regnault}},\ and\ \bibinfo {author} {\bibfnamefont {B.~A.}\ \bibnamefont {Bernevig}},\ }\bibfield  {title} {\bibinfo {title} {{Decomposition of fractional quantum Hall model states: Product rule symmetries and approximations}},\ }\href {https://doi.org/10.1103/PhysRevB.84.045127} {\bibfield  {journal} {\bibinfo  {journal} {Phys. Rev. B}\ }\textbf {\bibinfo {volume} {84}},\ \bibinfo {pages} {045127} (\bibinfo {year} {2011})}\BibitemShut {NoStop}%
\bibitem [{\citenamefont {Calogero}(1969{\natexlab{a}})}]{calogero69jmp2191}%
  \BibitemOpen
  \bibfield  {author} {\bibinfo {author} {\bibfnamefont {F.}~\bibnamefont {Calogero}},\ }\bibfield  {title} {\bibinfo {title} {{Solution of a Three‑Body Problem in One Dimension}},\ }\href {https://doi.org/10.1063/1.1664820} {\bibfield  {journal} {\bibinfo  {journal} {J.~Math. Phys.}\ }\textbf {\bibinfo {volume} {10}},\ \bibinfo {pages} {2191} (\bibinfo {year} {1969}{\natexlab{a}})}\BibitemShut {NoStop}%
\bibitem [{\citenamefont {Calogero}(1969{\natexlab{b}})}]{calogero69jmp2197}%
  \BibitemOpen
  \bibfield  {author} {\bibinfo {author} {\bibfnamefont {F.}~\bibnamefont {Calogero}},\ }\bibfield  {title} {\bibinfo {title} {{Ground State of a One‑Dimensional $N$‑Body System}},\ }\href {https://doi.org/10.1063/1.1664821} {\bibfield  {journal} {\bibinfo  {journal} {J.~Math. Phys.}\ }\textbf {\bibinfo {volume} {10}},\ \bibinfo {pages} {2197} (\bibinfo {year} {1969}{\natexlab{b}})}\BibitemShut {NoStop}%
\bibitem [{\citenamefont {Sutherland}(1971{\natexlab{a}})}]{sutherland71jmp246}%
  \BibitemOpen
  \bibfield  {author} {\bibinfo {author} {\bibfnamefont {B.}~\bibnamefont {Sutherland}},\ }\bibfield  {title} {\bibinfo {title} {{Quantum Many‑Body Problem in One Dimension: Ground State}},\ }\href {https://doi.org/10.1063/1.1665584} {\bibfield  {journal} {\bibinfo  {journal} {J.~Math. Phys.}\ }\textbf {\bibinfo {volume} {12}},\ \bibinfo {pages} {246} (\bibinfo {year} {1971}{\natexlab{a}})}\BibitemShut {NoStop}%
\bibitem [{\citenamefont {Sutherland}(1971{\natexlab{b}})}]{sutherland71pra2019}%
  \BibitemOpen
  \bibfield  {author} {\bibinfo {author} {\bibfnamefont {B.}~\bibnamefont {Sutherland}},\ }\bibfield  {title} {\bibinfo {title} {{Exact Results for a Quantum Many-Body Problem in One Dimension}},\ }\href {https://doi.org/10.1103/PhysRevA.4.2019} {\bibfield  {journal} {\bibinfo  {journal} {Phys. Rev. A}\ }\textbf {\bibinfo {volume} {4}},\ \bibinfo {pages} {2019} (\bibinfo {year} {1971}{\natexlab{b}})}\BibitemShut {NoStop}%
\bibitem [{\citenamefont {Sutherland}(1972)}]{sutherland72pra1372}%
  \BibitemOpen
  \bibfield  {author} {\bibinfo {author} {\bibfnamefont {B.}~\bibnamefont {Sutherland}},\ }\bibfield  {title} {\bibinfo {title} {{Exact Results for a Quantum Many-Body Problem in One Dimension. II}},\ }\href {https://doi.org/10.1103/PhysRevA.5.1372} {\bibfield  {journal} {\bibinfo  {journal} {Phys. Rev. A}\ }\textbf {\bibinfo {volume} {5}},\ \bibinfo {pages} {1372} (\bibinfo {year} {1972})}\BibitemShut {NoStop}%
\bibitem [{\citenamefont {Pasquier}(1994)}]{10.1007/3-540-58453-6_3}%
  \BibitemOpen
  \bibfield  {author} {\bibinfo {author} {\bibfnamefont {V.}~\bibnamefont {Pasquier}},\ }\bibfield  {title} {\bibinfo {title} {{A lecture on the Calogero-Sutherland models}},\ }in\ \href {https://doi.org/10.1007/3-540-58453-6_3} {\emph {\bibinfo {booktitle} {Integrable Models and Strings}}},\ \bibinfo {editor} {edited by\ \bibinfo {editor} {\bibfnamefont {A.}~\bibnamefont {Alekseev}}, \bibinfo {editor} {\bibfnamefont {A.}~\bibnamefont {Hietam{\"a}ki}}, \bibinfo {editor} {\bibfnamefont {K.}~\bibnamefont {Huitu}}, \bibinfo {editor} {\bibfnamefont {A.}~\bibnamefont {Morozov}},\ and\ \bibinfo {editor} {\bibfnamefont {A.}~\bibnamefont {Niemi}}}\ (\bibinfo  {publisher} {Springer},\ \bibinfo {address} {Berlin, Heidelberg},\ \bibinfo {year} {1994})\ pp.\ \bibinfo {pages} {36--48}\BibitemShut {NoStop}%
\bibitem [{\citenamefont {Ferrando}\ \emph {et~al.}(2025)\citenamefont {Ferrando}, \citenamefont {Lamers}, \citenamefont {Levkovich-Maslyuk},\ and\ \citenamefont {Serban}}]{ferrando2025bethe}%
  \BibitemOpen
  \bibfield  {author} {\bibinfo {author} {\bibfnamefont {G.}~\bibnamefont {Ferrando}}, \bibinfo {author} {\bibfnamefont {J.}~\bibnamefont {Lamers}}, \bibinfo {author} {\bibfnamefont {F.}~\bibnamefont {Levkovich-Maslyuk}},\ and\ \bibinfo {author} {\bibfnamefont {D.}~\bibnamefont {Serban}},\ }\bibfield  {title} {\bibinfo {title} {{Bethe Ansatz inside Calogero-Sutherland models}},\ }\href {https://doi.org/10.21468/SciPostPhys.18.1.035} {\bibfield  {journal} {\bibinfo  {journal} {SciPost Phys.}\ }\textbf {\bibinfo {volume} {18}},\ \bibinfo {pages} {035} (\bibinfo {year} {2025})}\BibitemShut {NoStop}%
\bibitem [{\citenamefont {Zirnbauer}\ and\ \citenamefont {Haldane}(1995)}]{PhysRevB.52.8729}%
  \BibitemOpen
  \bibfield  {author} {\bibinfo {author} {\bibfnamefont {M.~R.}\ \bibnamefont {Zirnbauer}}\ and\ \bibinfo {author} {\bibfnamefont {F.~D.~M.}\ \bibnamefont {Haldane}},\ }\bibfield  {title} {\bibinfo {title} {{Single-particle Green's functions of the Calogero-Sutherland model at couplings \ensuremath{\lambda}=1/2, 1, and 2}},\ }\href {https://doi.org/10.1103/PhysRevB.52.8729} {\bibfield  {journal} {\bibinfo  {journal} {Phys. Rev. B}\ }\textbf {\bibinfo {volume} {52}},\ \bibinfo {pages} {8729} (\bibinfo {year} {1995})}\BibitemShut {NoStop}%
\bibitem [{\citenamefont {Haldane}(1988)}]{haldane88prl635}%
  \BibitemOpen
  \bibfield  {author} {\bibinfo {author} {\bibfnamefont {F.~D.~M.}\ \bibnamefont {Haldane}},\ }\bibfield  {title} {\bibinfo {title} {{Exact Jastrow-Gutzwiller resonating-valence-bond ground state of the spin-$\frac{1}{2}$ antiferromagnetic Heisenberg chain with 1/${\mathrm{r}}^{2}$ exchange}},\ }\href {https://doi.org/10.1103/PhysRevLett.60.635} {\bibfield  {journal} {\bibinfo  {journal} {Phys. Rev. Lett.}\ }\textbf {\bibinfo {volume} {60}},\ \bibinfo {pages} {635} (\bibinfo {year} {1988})}\BibitemShut {NoStop}%
\bibitem [{\citenamefont {Shastry}(1988)}]{shastry88prl639}%
  \BibitemOpen
  \bibfield  {author} {\bibinfo {author} {\bibfnamefont {B.~S.}\ \bibnamefont {Shastry}},\ }\bibfield  {title} {\bibinfo {title} {{Exact solution of an $S=1/2$ Heisenberg antiferromagnetic chain with long-ranged interactions}},\ }\href {https://doi.org/10.1103/PhysRevLett.60.639} {\bibfield  {journal} {\bibinfo  {journal} {Phys. Rev. Lett.}\ }\textbf {\bibinfo {volume} {60}},\ \bibinfo {pages} {639} (\bibinfo {year} {1988})}\BibitemShut {NoStop}%
\bibitem [{\citenamefont {Polychronakos}(1993)}]{polychronakos93prl2329}%
  \BibitemOpen
  \bibfield  {author} {\bibinfo {author} {\bibfnamefont {A.~P.}\ \bibnamefont {Polychronakos}},\ }\bibfield  {title} {\bibinfo {title} {{Lattice integrable systems of Haldane-Shastry type}},\ }\href {https://doi.org/10.1103/PhysRevLett.70.2329} {\bibfield  {journal} {\bibinfo  {journal} {Phys. Rev. Lett.}\ }\textbf {\bibinfo {volume} {70}},\ \bibinfo {pages} {2329} (\bibinfo {year} {1993})}\BibitemShut {NoStop}%
\bibitem [{\citenamefont {Nielsen}\ \emph {et~al.}(2011)\citenamefont {Nielsen}, \citenamefont {Ignacio~Cirac},\ and\ \citenamefont {Sierra}}]{Nielsen_2011}%
  \BibitemOpen
  \bibfield  {author} {\bibinfo {author} {\bibfnamefont {A.~E.~B.}\ \bibnamefont {Nielsen}}, \bibinfo {author} {\bibfnamefont {J.}~\bibnamefont {Ignacio~Cirac}},\ and\ \bibinfo {author} {\bibfnamefont {G.}~\bibnamefont {Sierra}},\ }\bibfield  {title} {\bibinfo {title} {Quantum spin hamiltonians for the su(2)k wzw model},\ }\href {https://doi.org/10.1088/1742-5468/2011/11/P11014} {\bibfield  {journal} {\bibinfo  {journal} {Journal of Statistical Mechanics: Theory and Experiment}\ }\textbf {\bibinfo {volume} {2011}},\ \bibinfo {pages} {P11014} (\bibinfo {year} {2011})}\BibitemShut {NoStop}%
\bibitem [{\citenamefont {Thomale}\ \emph {et~al.}(2012)\citenamefont {Thomale}, \citenamefont {Rachel}, \citenamefont {Schmitteckert},\ and\ \citenamefont {Greiter}}]{PhysRevB.85.195149}%
  \BibitemOpen
  \bibfield  {author} {\bibinfo {author} {\bibfnamefont {R.}~\bibnamefont {Thomale}}, \bibinfo {author} {\bibfnamefont {S.}~\bibnamefont {Rachel}}, \bibinfo {author} {\bibfnamefont {P.}~\bibnamefont {Schmitteckert}},\ and\ \bibinfo {author} {\bibfnamefont {M.}~\bibnamefont {Greiter}},\ }\bibfield  {title} {\bibinfo {title} {{Family of spin-$S$ chain representations of SU(2)${}_{k}$ Wess-Zumino-Witten models}},\ }\href {https://doi.org/10.1103/PhysRevB.85.195149} {\bibfield  {journal} {\bibinfo  {journal} {Phys. Rev. B}\ }\textbf {\bibinfo {volume} {85}},\ \bibinfo {pages} {195149} (\bibinfo {year} {2012})}\BibitemShut {NoStop}%
\bibitem [{\citenamefont {Schroeter}\ \emph {et~al.}(2007)\citenamefont {Schroeter}, \citenamefont {Kapit}, \citenamefont {Thomale},\ and\ \citenamefont {Greiter}}]{PhysRevLett.99.097202}%
  \BibitemOpen
  \bibfield  {author} {\bibinfo {author} {\bibfnamefont {D.~F.}\ \bibnamefont {Schroeter}}, \bibinfo {author} {\bibfnamefont {E.}~\bibnamefont {Kapit}}, \bibinfo {author} {\bibfnamefont {R.}~\bibnamefont {Thomale}},\ and\ \bibinfo {author} {\bibfnamefont {M.}~\bibnamefont {Greiter}},\ }\bibfield  {title} {\bibinfo {title} {{Spin Hamiltonian for which the Chiral Spin Liquid is the Exact Ground State}},\ }\href {https://doi.org/10.1103/PhysRevLett.99.097202} {\bibfield  {journal} {\bibinfo  {journal} {Phys. Rev. Lett.}\ }\textbf {\bibinfo {volume} {99}},\ \bibinfo {pages} {097202} (\bibinfo {year} {2007})}\BibitemShut {NoStop}%
\bibitem [{\citenamefont {Thomale}\ \emph {et~al.}(2009)\citenamefont {Thomale}, \citenamefont {Kapit}, \citenamefont {Schroeter},\ and\ \citenamefont {Greiter}}]{PhysRevB.80.104406}%
  \BibitemOpen
  \bibfield  {author} {\bibinfo {author} {\bibfnamefont {R.}~\bibnamefont {Thomale}}, \bibinfo {author} {\bibfnamefont {E.}~\bibnamefont {Kapit}}, \bibinfo {author} {\bibfnamefont {D.~F.}\ \bibnamefont {Schroeter}},\ and\ \bibinfo {author} {\bibfnamefont {M.}~\bibnamefont {Greiter}},\ }\bibfield  {title} {\bibinfo {title} {{Parent Hamiltonian for the chiral spin liquid}},\ }\href {https://doi.org/10.1103/PhysRevB.80.104406} {\bibfield  {journal} {\bibinfo  {journal} {Phys. Rev. B}\ }\textbf {\bibinfo {volume} {80}},\ \bibinfo {pages} {104406} (\bibinfo {year} {2009})}\BibitemShut {NoStop}%
\bibitem [{\citenamefont {Greiter}\ and\ \citenamefont {Thomale}(2009)}]{PhysRevLett.102.207203}%
  \BibitemOpen
  \bibfield  {author} {\bibinfo {author} {\bibfnamefont {M.}~\bibnamefont {Greiter}}\ and\ \bibinfo {author} {\bibfnamefont {R.}~\bibnamefont {Thomale}},\ }\bibfield  {title} {\bibinfo {title} {{Non-Abelian Statistics in a Quantum Antiferromagnet}},\ }\href {https://doi.org/10.1103/PhysRevLett.102.207203} {\bibfield  {journal} {\bibinfo  {journal} {Phys. Rev. Lett.}\ }\textbf {\bibinfo {volume} {102}},\ \bibinfo {pages} {207203} (\bibinfo {year} {2009})}\BibitemShut {NoStop}%
\bibitem [{\citenamefont {Greiter}\ \emph {et~al.}(2014)\citenamefont {Greiter}, \citenamefont {Schroeter},\ and\ \citenamefont {Thomale}}]{PhysRevB.89.165125}%
  \BibitemOpen
  \bibfield  {author} {\bibinfo {author} {\bibfnamefont {M.}~\bibnamefont {Greiter}}, \bibinfo {author} {\bibfnamefont {D.~F.}\ \bibnamefont {Schroeter}},\ and\ \bibinfo {author} {\bibfnamefont {R.}~\bibnamefont {Thomale}},\ }\bibfield  {title} {\bibinfo {title} {{Parent Hamiltonian for the non-Abelian chiral spin liquid}},\ }\href {https://doi.org/10.1103/PhysRevB.89.165125} {\bibfield  {journal} {\bibinfo  {journal} {Phys. Rev. B}\ }\textbf {\bibinfo {volume} {89}},\ \bibinfo {pages} {165125} (\bibinfo {year} {2014})}\BibitemShut {NoStop}%
\bibitem [{\citenamefont {Haldane}\ \emph {et~al.}(1992)\citenamefont {Haldane}, \citenamefont {Ha}, \citenamefont {Talstra}, \citenamefont {Bernard},\ and\ \citenamefont {Pasquier}}]{PhysRevLett.69.2021}%
  \BibitemOpen
  \bibfield  {author} {\bibinfo {author} {\bibfnamefont {F.~D.~M.}\ \bibnamefont {Haldane}}, \bibinfo {author} {\bibfnamefont {Z.~N.~C.}\ \bibnamefont {Ha}}, \bibinfo {author} {\bibfnamefont {J.~C.}\ \bibnamefont {Talstra}}, \bibinfo {author} {\bibfnamefont {D.}~\bibnamefont {Bernard}},\ and\ \bibinfo {author} {\bibfnamefont {V.}~\bibnamefont {Pasquier}},\ }\bibfield  {title} {\bibinfo {title} {{Yangian symmetry of integrable quantum chains with long-range interactions and a new description of states in conformal field theory}},\ }\href {https://doi.org/10.1103/PhysRevLett.69.2021} {\bibfield  {journal} {\bibinfo  {journal} {Phys. Rev. Lett.}\ }\textbf {\bibinfo {volume} {69}},\ \bibinfo {pages} {2021} (\bibinfo {year} {1992})}\BibitemShut {NoStop}%
\bibitem [{\citenamefont {Talstra}\ and\ \citenamefont {Haldane}(1995)}]{JCTalstra_1995}%
  \BibitemOpen
  \bibfield  {author} {\bibinfo {author} {\bibfnamefont {J.~C.}\ \bibnamefont {Talstra}}\ and\ \bibinfo {author} {\bibfnamefont {F.~D.~M.}\ \bibnamefont {Haldane}},\ }\bibfield  {title} {\bibinfo {title} {{Integrals of motion of the Haldane-Shastry model}},\ }\href {https://doi.org/10.1088/0305-4470/28/8/027} {\bibfield  {journal} {\bibinfo  {journal} {J. Phys. A: Math. Gen.}\ }\textbf {\bibinfo {volume} {28}},\ \bibinfo {pages} {2369} (\bibinfo {year} {1995})}\BibitemShut {NoStop}%
\bibitem [{\citenamefont {Feiguin}\ \emph {et~al.}(2007)\citenamefont {Feiguin}, \citenamefont {Trebst}, \citenamefont {Ludwig}, \citenamefont {Troyer}, \citenamefont {Kitaev}, \citenamefont {Wang},\ and\ \citenamefont {Freedman}}]{PhysRevLett.98.160409}%
  \BibitemOpen
  \bibfield  {author} {\bibinfo {author} {\bibfnamefont {A.}~\bibnamefont {Feiguin}}, \bibinfo {author} {\bibfnamefont {S.}~\bibnamefont {Trebst}}, \bibinfo {author} {\bibfnamefont {A.~W.~W.}\ \bibnamefont {Ludwig}}, \bibinfo {author} {\bibfnamefont {M.}~\bibnamefont {Troyer}}, \bibinfo {author} {\bibfnamefont {A.}~\bibnamefont {Kitaev}}, \bibinfo {author} {\bibfnamefont {Z.}~\bibnamefont {Wang}},\ and\ \bibinfo {author} {\bibfnamefont {M.~H.}\ \bibnamefont {Freedman}},\ }\bibfield  {title} {\bibinfo {title} {{Interacting Anyons in Topological Quantum Liquids: The Golden Chain}},\ }\href {https://doi.org/10.1103/PhysRevLett.98.160409} {\bibfield  {journal} {\bibinfo  {journal} {Phys. Rev. Lett.}\ }\textbf {\bibinfo {volume} {98}},\ \bibinfo {pages} {160409} (\bibinfo {year} {2007})}\BibitemShut {NoStop}%
\bibitem [{\citenamefont {{Belavin}}\ \emph {et~al.}(1984)\citenamefont {{Belavin}}, \citenamefont {{Polyakov}},\ and\ \citenamefont {{Zamolodchikov}}}]{1984NuPhB.241..333B}%
  \BibitemOpen
  \bibfield  {author} {\bibinfo {author} {\bibfnamefont {A.~A.}\ \bibnamefont {{Belavin}}}, \bibinfo {author} {\bibfnamefont {A.~M.}\ \bibnamefont {{Polyakov}}},\ and\ \bibinfo {author} {\bibfnamefont {A.~B.}\ \bibnamefont {{Zamolodchikov}}},\ }\bibfield  {title} {\bibinfo {title} {{Infinite conformal symmetry in two-dimensional quantum field theory}},\ }\href {https://doi.org/10.1016/0550-3213(84)90052-X} {\bibfield  {journal} {\bibinfo  {journal} {Nucl. Phys. B}\ }\textbf {\bibinfo {volume} {241}},\ \bibinfo {pages} {333} (\bibinfo {year} {1984})}\BibitemShut {NoStop}%
\bibitem [{\citenamefont {Ginsparg}(1988)}]{ginsparg_1988}%
  \BibitemOpen
  \bibfield  {author} {\bibinfo {author} {\bibfnamefont {P.}~\bibnamefont {Ginsparg}},\ }\href {https://arxiv.org/abs/hep-th/9108028} {\bibinfo {title} {{Applied Conformal Field Theory}}} (\bibinfo {year} {1988}),\ \Eprint {https://arxiv.org/abs/hep-th/9108028} {arXiv:hep-th/9108028 [hep-th]} \BibitemShut {NoStop}%
\bibitem [{\citenamefont {Di~Francesco}\ \emph {et~al.}(1997)\citenamefont {Di~Francesco}, \citenamefont {Mathieu},\ and\ \citenamefont {Sénéchal}}]{DiFrancesco_1997}%
  \BibitemOpen
  \bibfield  {author} {\bibinfo {author} {\bibfnamefont {P.}~\bibnamefont {Di~Francesco}}, \bibinfo {author} {\bibfnamefont {P.}~\bibnamefont {Mathieu}},\ and\ \bibinfo {author} {\bibfnamefont {D.}~\bibnamefont {Sénéchal}},\ }\href {https://doi.org/10.1007/978-1-4612-2256-9} {\emph {\bibinfo {title} {{Conformal Field Theory}}}},\ Graduate Texts in Contemporary Physics\ (\bibinfo  {publisher} {Springer-Verlag},\ \bibinfo {address} {New York},\ \bibinfo {year} {1997})\BibitemShut {NoStop}%
\bibitem [{\citenamefont {Laughlin}(1983)}]{PhysRevLett.50.1395}%
  \BibitemOpen
  \bibfield  {author} {\bibinfo {author} {\bibfnamefont {R.~B.}\ \bibnamefont {Laughlin}},\ }\bibfield  {title} {\bibinfo {title} {Anomalous quantum hall effect: An incompressible quantum fluid with fractionally charged excitations},\ }\href {https://doi.org/10.1103/PhysRevLett.50.1395} {\bibfield  {journal} {\bibinfo  {journal} {Phys. Rev. Lett.}\ }\textbf {\bibinfo {volume} {50}},\ \bibinfo {pages} {1395} (\bibinfo {year} {1983})}\BibitemShut {NoStop}%
\bibitem [{\citenamefont {Awata}\ \emph {et~al.}(1995)\citenamefont {Awata}, \citenamefont {Matsuo}, \citenamefont {Odake},\ and\ \citenamefont {Shiraishi}}]{AWATA1995347}%
  \BibitemOpen
  \bibfield  {author} {\bibinfo {author} {\bibfnamefont {H.}~\bibnamefont {Awata}}, \bibinfo {author} {\bibfnamefont {Y.}~\bibnamefont {Matsuo}}, \bibinfo {author} {\bibfnamefont {S.}~\bibnamefont {Odake}},\ and\ \bibinfo {author} {\bibfnamefont {J.}~\bibnamefont {Shiraishi}},\ }\bibfield  {title} {\bibinfo {title} {{Excited states of the Calogero-Sutherland model and singular vectors of the WN algebra}},\ }\href {https://doi.org/https://doi.org/10.1016/0550-3213(95)00286-2} {\bibfield  {journal} {\bibinfo  {journal} {Nucl. Phys. B}\ }\textbf {\bibinfo {volume} {449}},\ \bibinfo {pages} {347} (\bibinfo {year} {1995})}\BibitemShut {NoStop}%
\bibitem [{\citenamefont {Lapointe}\ and\ \citenamefont {Vinet}(1996)}]{lapointevinet}%
  \BibitemOpen
  \bibfield  {author} {\bibinfo {author} {\bibfnamefont {L.}~\bibnamefont {Lapointe}}\ and\ \bibinfo {author} {\bibfnamefont {L.}~\bibnamefont {Vinet}},\ }\bibfield  {title} {\bibinfo {title} {{Exact operator solution of the Calogero-Sutherland model}},\ }\href {https://doi.org/10.1007/BF02099456} {\bibfield  {journal} {\bibinfo  {journal} {Commun. Math. Phys.}\ }\textbf {\bibinfo {volume} {178}},\ \bibinfo {pages} {425} (\bibinfo {year} {1996})}\BibitemShut {NoStop}%
\bibitem [{\citenamefont {Bernevig}\ and\ \citenamefont {Haldane}(2008)}]{Bernevig_2008}%
  \BibitemOpen
  \bibfield  {author} {\bibinfo {author} {\bibfnamefont {B.~A.}\ \bibnamefont {Bernevig}}\ and\ \bibinfo {author} {\bibfnamefont {F.~D.~M.}\ \bibnamefont {Haldane}},\ }\bibfield  {title} {\bibinfo {title} {{Model Fractional Quantum Hall States and Jack Polynomials}},\ }\href {https://doi.org/10.1103/PhysRevLett.100.246802} {\bibfield  {journal} {\bibinfo  {journal} {Phys. Rev. Lett.}\ }\textbf {\bibinfo {volume} {100}},\ \bibinfo {pages} {246802} (\bibinfo {year} {2008})}\BibitemShut {NoStop}%
\bibitem [{\citenamefont {del Campo}(2020)}]{Adolfo-2020}%
  \BibitemOpen
  \bibfield  {author} {\bibinfo {author} {\bibfnamefont {A.}~\bibnamefont {del Campo}},\ }\bibfield  {title} {\bibinfo {title} {{Exact ground states of quantum many-body systems under confinement}},\ }\href {https://doi.org/10.1103/PhysRevResearch.2.043114} {\bibfield  {journal} {\bibinfo  {journal} {Phys. Rev. Res.}\ }\textbf {\bibinfo {volume} {2}},\ \bibinfo {pages} {043114} (\bibinfo {year} {2020})}\BibitemShut {NoStop}%
\bibitem [{\citenamefont {Beau}\ and\ \citenamefont {del Campo}(2021)}]{Beau-2021}%
  \BibitemOpen
  \bibfield  {author} {\bibinfo {author} {\bibfnamefont {M.}~\bibnamefont {Beau}}\ and\ \bibinfo {author} {\bibfnamefont {A.}~\bibnamefont {del Campo}},\ }\bibfield  {title} {\bibinfo {title} {{Parent Hamiltonians of Jastrow wavefunctions}},\ }\href {https://doi.org/10.21468/SciPostPhysCore.4.4.030} {\bibfield  {journal} {\bibinfo  {journal} {SciPost Phys. Core}\ }\textbf {\bibinfo {volume} {4}},\ \bibinfo {pages} {030} (\bibinfo {year} {2021})}\BibitemShut {NoStop}%
\bibitem [{\citenamefont {Sasmal}\ and\ \citenamefont {del Campo}(2026)}]{Sasmal-2026}%
  \BibitemOpen
  \bibfield  {author} {\bibinfo {author} {\bibfnamefont {N.}~\bibnamefont {Sasmal}}\ and\ \bibinfo {author} {\bibfnamefont {A.}~\bibnamefont {del Campo}},\ }\bibfield  {title} {\bibinfo {title} {Taxonomy of integrable and ground-state solvable models: Jastrow wave functions on graphs and parent hamiltonians},\ }\href {https://doi.org/10.1103/g7hk-tlq1} {\bibfield  {journal} {\bibinfo  {journal} {Phys. Rev. B}\ }\textbf {\bibinfo {volume} {113}},\ \bibinfo {pages} {235151} (\bibinfo {year} {2026})}\BibitemShut {NoStop}%
\bibitem [{\citenamefont {Polychronakos}(1992)}]{Polychronakos-1992}%
  \BibitemOpen
  \bibfield  {author} {\bibinfo {author} {\bibfnamefont {A.~P.}\ \bibnamefont {Polychronakos}},\ }\bibfield  {title} {\bibinfo {title} {{Exchange operator formalism for integrable systems of particles}},\ }\href {https://doi.org/10.1103/PhysRevLett.69.703} {\bibfield  {journal} {\bibinfo  {journal} {Phys. Rev. Lett.}\ }\textbf {\bibinfo {volume} {69}},\ \bibinfo {pages} {703} (\bibinfo {year} {1992})}\BibitemShut {NoStop}%
\bibitem [{\citenamefont {Brink}\ \emph {et~al.}(1992)\citenamefont {Brink}, \citenamefont {Hansson},\ and\ \citenamefont {Vasiliev}}]{Brink_1992}%
  \BibitemOpen
  \bibfield  {author} {\bibinfo {author} {\bibfnamefont {L.}~\bibnamefont {Brink}}, \bibinfo {author} {\bibfnamefont {T.}~\bibnamefont {Hansson}},\ and\ \bibinfo {author} {\bibfnamefont {M.}~\bibnamefont {Vasiliev}},\ }\bibfield  {title} {\bibinfo {title} {Explicit solution to the n-body calogero problem},\ }\href {https://doi.org/https://doi.org/10.1016/0370-2693(92)90166-2} {\bibfield  {journal} {\bibinfo  {journal} {Physics Letters B}\ }\textbf {\bibinfo {volume} {286}},\ \bibinfo {pages} {109} (\bibinfo {year} {1992})}\BibitemShut {NoStop}%
\bibitem [{\citenamefont {Read}\ and\ \citenamefont {Rezayi}(1999)}]{Read-99prb8084}%
  \BibitemOpen
  \bibfield  {author} {\bibinfo {author} {\bibfnamefont {N.}~\bibnamefont {Read}}\ and\ \bibinfo {author} {\bibfnamefont {E.}~\bibnamefont {Rezayi}},\ }\bibfield  {title} {\bibinfo {title} {{Beyond paired quantum Hall states: Parafermions and incompressible states in the first excited Landau level}},\ }\href {https://doi.org/10.1103/PhysRevB.59.8084} {\bibfield  {journal} {\bibinfo  {journal} {Phys. Rev. B}\ }\textbf {\bibinfo {volume} {59}},\ \bibinfo {pages} {8084} (\bibinfo {year} {1999})}\BibitemShut {NoStop}%
\bibitem [{\citenamefont {Greiter}\ \emph {et~al.}(1991)\citenamefont {Greiter}, \citenamefont {Wen},\ and\ \citenamefont {Wilczek}}]{greiter-91prl3205}%
  \BibitemOpen
  \bibfield  {author} {\bibinfo {author} {\bibfnamefont {M.}~\bibnamefont {Greiter}}, \bibinfo {author} {\bibfnamefont {X.-G.}\ \bibnamefont {Wen}},\ and\ \bibinfo {author} {\bibfnamefont {F.}~\bibnamefont {Wilczek}},\ }\bibfield  {title} {\bibinfo {title} {{Paired Hall state at half filling}},\ }\href {https://doi.org/10.1103/PhysRevLett.66.3205} {\bibfield  {journal} {\bibinfo  {journal} {Phys. Rev. Lett.}\ }\textbf {\bibinfo {volume} {66}},\ \bibinfo {pages} {3205} (\bibinfo {year} {1991})}\BibitemShut {NoStop}%
\bibitem [{\citenamefont {Greiter}\ \emph {et~al.}(1992)\citenamefont {Greiter}, \citenamefont {Wen},\ and\ \citenamefont {Wilczek}}]{greiter-92prb9586}%
  \BibitemOpen
  \bibfield  {author} {\bibinfo {author} {\bibfnamefont {M.}~\bibnamefont {Greiter}}, \bibinfo {author} {\bibfnamefont {X.~G.}\ \bibnamefont {Wen}},\ and\ \bibinfo {author} {\bibfnamefont {F.}~\bibnamefont {Wilczek}},\ }\bibfield  {title} {\bibinfo {title} {{Paired Hall states in double-layer electron systems}},\ }\href {https://doi.org/10.1103/PhysRevB.46.9586} {\bibfield  {journal} {\bibinfo  {journal} {Phys. Rev. B}\ }\textbf {\bibinfo {volume} {46}},\ \bibinfo {pages} {9586} (\bibinfo {year} {1992})}\BibitemShut {NoStop}%
\bibitem [{\citenamefont {Read}\ and\ \citenamefont {Green}(2000)}]{Read_00}%
  \BibitemOpen
  \bibfield  {author} {\bibinfo {author} {\bibfnamefont {N.}~\bibnamefont {Read}}\ and\ \bibinfo {author} {\bibfnamefont {D.}~\bibnamefont {Green}},\ }\bibfield  {title} {\bibinfo {title} {{Paired states of fermions in two dimensions with breaking of parity and time-reversal symmetries and the fractional quantum Hall effect}},\ }\href {https://doi.org/10.1103/PhysRevB.61.10267} {\bibfield  {journal} {\bibinfo  {journal} {Phys. Rev. B}\ }\textbf {\bibinfo {volume} {61}},\ \bibinfo {pages} {10267} (\bibinfo {year} {2000})}\BibitemShut {NoStop}%
\bibitem [{\citenamefont {Rezayi}\ and\ \citenamefont {Haldane}(2000)}]{Rezayi-00prl4685}%
  \BibitemOpen
  \bibfield  {author} {\bibinfo {author} {\bibfnamefont {E.~H.}\ \bibnamefont {Rezayi}}\ and\ \bibinfo {author} {\bibfnamefont {F.~D.~M.}\ \bibnamefont {Haldane}},\ }\bibfield  {title} {\bibinfo {title} {{Incompressible Paired Hall State, Stripe Order, and the Composite Fermion Liquid Phase in Half-Filled Landau Levels}},\ }\href {https://doi.org/10.1103/PhysRevLett.84.4685} {\bibfield  {journal} {\bibinfo  {journal} {Phys. Rev. Lett.}\ }\textbf {\bibinfo {volume} {84}},\ \bibinfo {pages} {4685} (\bibinfo {year} {2000})}\BibitemShut {NoStop}%
\bibitem [{\citenamefont {Banerjee}\ \emph {et~al.}(2018)\citenamefont {Banerjee}, \citenamefont {Heiblum}, \citenamefont {Umansky}, \citenamefont {Feldman}, \citenamefont {Oreg},\ and\ \citenamefont {Stern}}]{Banerjee2018}%
  \BibitemOpen
  \bibfield  {author} {\bibinfo {author} {\bibfnamefont {M.}~\bibnamefont {Banerjee}}, \bibinfo {author} {\bibfnamefont {M.}~\bibnamefont {Heiblum}}, \bibinfo {author} {\bibfnamefont {V.}~\bibnamefont {Umansky}}, \bibinfo {author} {\bibfnamefont {D.~E.}\ \bibnamefont {Feldman}}, \bibinfo {author} {\bibfnamefont {Y.}~\bibnamefont {Oreg}},\ and\ \bibinfo {author} {\bibfnamefont {A.}~\bibnamefont {Stern}},\ }\bibfield  {title} {\bibinfo {title} {{Observation of half-integer thermal Hall conductance}},\ }\href {https://doi.org/10.1038/s41586-018-0184-1} {\bibfield  {journal} {\bibinfo  {journal} {Nature (London)}\ }\textbf {\bibinfo {volume} {559}},\ \bibinfo {pages} {205} (\bibinfo {year} {2018})}\BibitemShut {NoStop}%
\bibitem [{\citenamefont {Lotrič}\ \emph {et~al.}(2025)\citenamefont {Lotrič}, \citenamefont {Wang}, \citenamefont {Zaletel}, \citenamefont {Simon},\ and\ \citenamefont {Parameswaran}}]{lotric_2025}%
  \BibitemOpen
  \bibfield  {author} {\bibinfo {author} {\bibfnamefont {T.}~\bibnamefont {Lotrič}}, \bibinfo {author} {\bibfnamefont {T.}~\bibnamefont {Wang}}, \bibinfo {author} {\bibfnamefont {M.~P.}\ \bibnamefont {Zaletel}}, \bibinfo {author} {\bibfnamefont {S.~H.}\ \bibnamefont {Simon}},\ and\ \bibinfo {author} {\bibfnamefont {S.~A.}\ \bibnamefont {Parameswaran}},\ }\href {https://arxiv.org/abs/2507.07161} {\bibinfo {title} {{Majorana edge reconstruction and the $\nu=5/2$ non-Abelian thermal Hall puzzle}}} (\bibinfo {year} {2025}),\ \Eprint {https://arxiv.org/abs/2507.07161} {arXiv:2507.07161 [cond-mat.mes-hall]} \BibitemShut {NoStop}%
\bibitem [{\citenamefont {Xia}\ \emph {et~al.}(2004)\citenamefont {Xia}, \citenamefont {Pan}, \citenamefont {Vicente}, \citenamefont {Adams}, \citenamefont {Sullivan}, \citenamefont {Stormer}, \citenamefont {Tsui}, \citenamefont {Pfeiffer}, \citenamefont {Baldwin},\ and\ \citenamefont {West}}]{xia-04prl176809}%
  \BibitemOpen
  \bibfield  {author} {\bibinfo {author} {\bibfnamefont {J.~S.}\ \bibnamefont {Xia}}, \bibinfo {author} {\bibfnamefont {W.}~\bibnamefont {Pan}}, \bibinfo {author} {\bibfnamefont {C.~L.}\ \bibnamefont {Vicente}}, \bibinfo {author} {\bibfnamefont {E.~D.}\ \bibnamefont {Adams}}, \bibinfo {author} {\bibfnamefont {N.~S.}\ \bibnamefont {Sullivan}}, \bibinfo {author} {\bibfnamefont {H.~L.}\ \bibnamefont {Stormer}}, \bibinfo {author} {\bibfnamefont {D.~C.}\ \bibnamefont {Tsui}}, \bibinfo {author} {\bibfnamefont {L.~N.}\ \bibnamefont {Pfeiffer}}, \bibinfo {author} {\bibfnamefont {K.~W.}\ \bibnamefont {Baldwin}},\ and\ \bibinfo {author} {\bibfnamefont {K.~W.}\ \bibnamefont {West}},\ }\bibfield  {title} {\bibinfo {title} {{Electron Correlation in the Second Landau Level: A Competition Between Many Nearly Degenerate Quantum Phases}},\ }\href {https://doi.org/10.1103/PhysRevLett.93.176809} {\bibfield  {journal} {\bibinfo  {journal} {Phys. Rev. Lett.}\ }\textbf {\bibinfo {volume} {93}},\ \bibinfo {pages} {176809} (\bibinfo
  {year} {2004})}\BibitemShut {NoStop}%
\bibitem [{\citenamefont {Rezayi}\ and\ \citenamefont {Read}(2009)}]{Rezayi_2009}%
  \BibitemOpen
  \bibfield  {author} {\bibinfo {author} {\bibfnamefont {E.~H.}\ \bibnamefont {Rezayi}}\ and\ \bibinfo {author} {\bibfnamefont {N.}~\bibnamefont {Read}},\ }\bibfield  {title} {\bibinfo {title} {{Non-Abelian quantized Hall states of electrons at filling factors $12/5$ and $13/5$ in the first excited Landau level}},\ }\href {https://doi.org/10.1103/PhysRevB.79.075306} {\bibfield  {journal} {\bibinfo  {journal} {Phys. Rev. B}\ }\textbf {\bibinfo {volume} {79}},\ \bibinfo {pages} {075306} (\bibinfo {year} {2009})}\BibitemShut {NoStop}%
\bibitem [{\citenamefont {Thomale}\ \emph {et~al.}(2010{\natexlab{a}})\citenamefont {Thomale}, \citenamefont {Sterdyniak}, \citenamefont {Regnault},\ and\ \citenamefont {Bernevig}}]{Thomale_2010}%
  \BibitemOpen
  \bibfield  {author} {\bibinfo {author} {\bibfnamefont {R.}~\bibnamefont {Thomale}}, \bibinfo {author} {\bibfnamefont {A.}~\bibnamefont {Sterdyniak}}, \bibinfo {author} {\bibfnamefont {N.}~\bibnamefont {Regnault}},\ and\ \bibinfo {author} {\bibfnamefont {B.~A.}\ \bibnamefont {Bernevig}},\ }\bibfield  {title} {\bibinfo {title} {Entanglement gap and a new principle of adiabatic continuity},\ }\href {https://doi.org/10.1103/PhysRevLett.104.180502} {\bibfield  {journal} {\bibinfo  {journal} {Phys. Rev. Lett.}\ }\textbf {\bibinfo {volume} {104}},\ \bibinfo {pages} {180502} (\bibinfo {year} {2010}{\natexlab{a}})}\BibitemShut {NoStop}%
\bibitem [{\citenamefont {Thomale}\ \emph {et~al.}(2010{\natexlab{b}})\citenamefont {Thomale}, \citenamefont {Arovas},\ and\ \citenamefont {Bernevig}}]{Thomale_2011}%
  \BibitemOpen
  \bibfield  {author} {\bibinfo {author} {\bibfnamefont {R.}~\bibnamefont {Thomale}}, \bibinfo {author} {\bibfnamefont {D.~P.}\ \bibnamefont {Arovas}},\ and\ \bibinfo {author} {\bibfnamefont {B.~A.}\ \bibnamefont {Bernevig}},\ }\bibfield  {title} {\bibinfo {title} {Nonlocal order in gapless systems: Entanglement spectrum in spin chains},\ }\href {https://doi.org/10.1103/PhysRevLett.105.116805} {\bibfield  {journal} {\bibinfo  {journal} {Phys. Rev. Lett.}\ }\textbf {\bibinfo {volume} {105}},\ \bibinfo {pages} {116805} (\bibinfo {year} {2010}{\natexlab{b}})}\BibitemShut {NoStop}%
\end{thebibliography}

%

\end{document}